\documentclass[prd,aps,twocolumn,nofootinbib,superscriptaddress,eqsecnum,floatfix,preprintnumbers,amsmath,amssymb,
nofootinbib,longbibliography,]{revtex4-1}

\usepackage{amssymb,amsmath}
\usepackage{epsfig}
\usepackage[dvipsnames]{xcolor}
\usepackage[utf8]{inputenc}
\usepackage{stmaryrd}
\usepackage{mathrsfs}
\usepackage{mathalfa}
\usepackage{accents}
\usepackage[normalem]{ulem}

\newcommand{\dd}{\mathrm{d}}

\newcommand{\be}{\begin{equation}}
\newcommand{\ee}{\end{equation}}
\newcommand{\bea}{\begin{eqnarray}}
\newcommand{\eea}{\end{eqnarray}}
\newcommand{\ba}{\begin{equation}\begin{aligned}}
\newcommand{\ea}{\end{aligned}\end{equation}}

\newcommand{\beg}{\begin{gather*}}
\newcommand{\eng}{\end{gather*}}
\newcommand{\hh}{,\hspace{0.5cm}}
\newcommand{\hhh}{,\hspace{0.2cm}}

\newcommand{\eq}[1]{(\ref{#1})}

\newcommand{\n}[1]{\label{#1}}

\newcommand{\ins}[1]{{\mbox{\tiny #1}}}
\newcommand{\ind}[1]{{\mbox{\scriptsize #1}}}

\newcommand{\ts}[1]{{\boldsymbol{#1}}}

\def\XXint#1#2#3{{\setbox0=\hbox{$#1{#2#3}{\int}$ }
\vcenter{\hbox{$#2#3$ }}\kern-.6\wd0}}

\usepackage[makeroom]{cancel}
\usepackage[caption=false]{subfig}
\usepackage[colorlinks=true,
            citecolor=green,
            linkcolor=red,
            filecolor=cyan,
            urlcolor=magenta,
            backref=false]{hyperref}

\newcommand{\lp}{\lambda_+}
\newcommand{\lm}{\lambda_-}
\newcommand{\bp}{\bar{p}}
\newcommand{\bq}{\bar{q}}
\newcommand{\bc}{\bar{\chi}}

\begin{document}

\title{Bouncing cosmology in the limiting curvature theory of gravity}

\author{Valeri P. Frolov}
\email{vfrolov@ualberta.ca}
\affiliation{Theoretical Physics Institute, University of Alberta, Edmonton, Alberta, Canada T6G 2E1}
\author{Andrei Zelnikov}
\email{zelnikov@ualberta.ca}
\affiliation{Theoretical Physics Institute, University of Alberta, Edmonton, Alberta, Canada T6G 2E1}


\begin{abstract}
In this paper we discuss models satisfying the limiting curvature condition. For this purpose we modify the Einstein-Hilbert action by adding a term which restricts the growth of curvature. We analyze cosmological solutions in such models. Namely, we consider a closed contracting homogeneous isotropic universe filled with thermal radiation. We demonstrate that for properly chosen curvature constraints such a universe has a bounce. As a result its evolution is nonsingular and contains a ``de Sitter--type" supercritical stage connecting contracting and expanding phases. Possible generalizations of these results are briefly discussed.

\hfill {\scriptsize Alberta Thy 27-21}
\end{abstract}

\maketitle
\section{Introduction}

The idea that the Universe can have a prehistory before the big bang is very old. Cyclic or oscillating cosmological models were considered almost 90 years ago. Such models were discussed in the famous book by Tolman published in 1934 \cite{Tolman1987Relativity}. He also demonstrated that the validity of the second law of thermodynamics applied to the Universe and increasing entropy make pure periodic models impossible: each of the successive cycles should be longer and larger than the previous one.

Even if one does not require the existence of an infinite number of cycles before the formation of the present Universe it is interesting to analyze an option that the Universe before the big bang had a phase of contraction, usually called a big crunch. In such models the Universe should experience a bounce, where its size takes some minimal value. We denote the scale factor that enters the Friedmann-Robertson-Walker metric for a homogeneous isotropic universe as $a(t)$. Then, one of the Einstein equations implies that
\be
{\ddot{a}\over a}=-{4\pi G\over 3}(\varepsilon+3P)\, ,
\ee
where $\varepsilon$ and $P$ are the matter energy density and pressure, respectively.
Since at the bounce point where $\dot{a}=0$ one has $\ddot{a}>0$ the equation of state should be such that $\varepsilon+3P<0$. The famous Penrose-Hawking singularity theorems  imply that in a general case in order to escape a cosmological singularity some of the energy conditions for matter should be violated \cite{Hawking:1973uf}.

Singularities in standard cosmological models are connected with an infinite growth of the spacetime curvature. Markov \cite{Markov:1982, Markov:1984ii} suggested that the existence of the limiting curvature should be considered as a new physical principle. He demonstrated that for a proper choice of the equation of state in the cosmology the limiting curvature condition is satisfied and solutions in such a model describe a bouncing universe.
A bouncing universe was discussed by Gasperini and Veneziano in their pre-big-bang string cosmology \cite{Gasperini:1992em,Gasperini:2002bn}.
Nonsingular cosmological models that are based on the use of nondynamical scalar fields to implement the limiting curvature hypothesis were studied some time ago by Brandenberger, et al. \cite{Mukhanov:1991zn,Brandenberger:1993ef}.
More recently, the interest in bouncing cosmological models has increased. This is mainly connected with the remarkable increase in the accuracy of cosmological observations. An interesting and intriguing question is: if there was of a big crunch phase, is it possible to find observational evidence of this?
A variety of different proposed  bouncing cosmological models have been discussed in several nice review articles, which also contain references to the original publications \cite{Turok_2005,Biswas_2006,Barvinsky:2008ia,Novello:2008ra,Lehners:2008vx,Ashtekar_2009,CesareSilva:2020ihf,Biswas:2012bp,Battefeld:2014uga,Brandenberger_2017,Yoshida:2017swb,Ijjas_2018}.

In this paper we discuss bouncing cosmological models in a new recently proposed limiting curvature gravity (LCG) theory \cite{Frolov:2021kcv}. The main idea of this approach is to modify the Einstein-Hilbert action by adding a constraint term which controls the curvature behavior and forbids its infinite growth. In fact, this is a realization of the Markov's old idea about the existence of a limiting curvature. A limiting curvature modification of a two-dimensional  dilaton gravity was considered in  \cite{Frolov:2021kcv}. In this paper we discuss four-dimensional LCG models.  In the Friedmann-Robertson-Walker metric for a homogeneous isotropic universe the Weyl tensor vanishes. Therefore, it is sufficient to restrict the growth of the Ricci tensor.

We shall discuss two types of models. We first introduce linear-in-curvature constraints. For this purpose we add to the action terms that are linear in the Ricci scalar and the eigenvalues of the Ricci tensor. After this, we discuss quadratic-in-curvature constraints. In both cases, we demonstrate that there exists a wide class of curvature constraints for which the curvature remains uniformly bounded during the evolution of the universe. A common property of such limiting curvature gravity models is that the cosmological solutions have a bounce. A contracting universe at some stage of its evolution, when its curvature reaches the critical value, enters a supercritical regime. If the initial size of the universe was large, then the corresponding supercritical solution is always close to the de Sitter solution. After passing the bounce point the universe expands. We demonstrate that at some moment of time it can leave its supercritical regime and one gets an expanding universe filled with matter. After this it follows the standard Einstein equations.

The paper is organized as follows. In Sec.\,\ref{Sec2} we recall some well-known properties of the isotropic homogeneous cosmological models and introduce notations that are used later in the paper.  LCG models and reduced actions for these models are discussed in Sec.\,\ref{Sec3}. Sections \ref{Sec4}--\ref{Sec7} discuss LCG models with linear-in-curvature constraints. Sections \ref{Sec8}--\ref{Sec10} are devoted to study LCG models with quadratic-in-curvature constraints. More general curvature constraints are discussed in Sec.\,\ref{Sec11}. Finally, Sec\, \ref{Sec12} contains a summary of the obtained results, a discussion of different aspects of LCG cosmological models and their possible generalizations. Some additional technical details and results used in the main text are collected in the Appendix.

\section{Isotropic homogeneous cosmology}\label{Sec2}

Let us consider the cosmological metric in the form
\be \label{metr}
ds^2=-b^2(t) dt^2+a^2(t) d\gamma^2 \, .
\ee
This metric  is  a direct sum of the one-dimensional metric $b^2(t)dt^2$ and three-dimensional metric $a^2(t) d\gamma^2$, where $d\gamma^2=\gamma_{ij}dx^idx^j$ is a line element on a unit 3D sphere $S^3$.
The metric $d\gamma^2$ admits group $O(4)$ of symmetries. It is well known that:
\begin{itemize}
\item A scalar function on $S^3$ invariant under the action of this group is a constant.
\item There does not exist a nonvanishing vector field invariant under the group of symmetries.
\item A symmetric rank-two tensor field $A_{ij}$ on $S^3$ invariant under the group of symmetries is $A_{ij}=A\gamma_{ij}$, where $A$ is a constant.
\end{itemize}

Consider a symmetric tensor $\ts{A}$ in a spacetime with metric (\ref{metr}) which respects its symmetry. Then, it has the following form:
\be \n{AA}
A^{\mu}_{\nu}=\mbox{diag}({\cal A}(t),{\cal \hat{A}}(t),{\cal \hat{A}}(t),{\cal \hat{A}}(t))\, .
\ee
It is easy to see that ${\cal A}(t)$ and ${\cal \hat{A}}(t)$ are eigenvalues of the tensor $A^{\mu}_{\nu}$.
We call them temporal and spatial eigenvalues, respectively.

In what follows we use similar notations for other symmetric rank-two tensors. For example, the Ricci tensor $R_{\mu\nu}$ has the form
\be
R^{\mu}_{\nu}=\mbox{diag}({\cal R},{\cal \hat{R}},{\cal \hat{R}},{\cal \hat{R}})\, .
\ee
Then, the Ricci scalar is
\be\n{RRR}
R={\cal R}+3{\cal \hat{R}}.
\ee
We keep the coefficient $b^2(t)$ of the metric \eq{metr} as an arbitrary function. This will allow us to obtain a complete set of the gravitational field equations from a reduced metric, but later, after the variations, we put $b(t)=1$. This is nothing but a gauge-fixing condition corresponding to synchronous gauge.
The eigenvalues ${\cal R}$ and ${\cal \hat{R}}$ of the Ricci tensor can be expressed in terms of  two structures $p$ and $q$
\ba\n{Rpq}
&R=6(q+p)\hh {\cal R}=3q\hh {\cal \hat{R}}=q+2p   ,
\ea
\ba
&q=\frac{1}{b^2}\Big(\frac{\ddot{a}}{a}-\frac{\dot{a}}{a}\frac{\dot{b}}{b}\Big)\hh
p=\frac{\dot{a}^2+b^2}{a^2b^2} .\n{ppqq}
\ea
The traceless part of the Ricci tensor
\ba
S^{\mu}_{\nu}=R^{\mu}_{\nu}-\frac{1}{4}\delta^{\mu}_{\nu}R
\ea
has the following eigenvalues
\be
S^{\mu}_{\nu}=\mbox{diag}({\cal S},{\cal \hat{S}},{\cal \hat{S}},{\cal \hat{S}})\, ,
\ee
where
\bea\n{Spq}
&{\cal S}=\frac{3}{2}(q-p)\hh {\cal \hat{S}}=-\frac{1}{2}(q-p) .
\eea

The Weyl tensor for the metric (\ref{metr}) vanishes, that is, all of the information about the spacetime curvature is encoded in the Ricci tensor. Our goal is to study cosmological models that obey the limiting curvature condition. A natural way to do this is to impose restrictions on the eigenvalues of the Ricci tensor. For example, one may try to restrict the value of the Ricci scalar (\ref{RRR}) which is a linear combination of these eigenvalues. However, the form (\ref{Rpq}) of this invariant implies that this does not work. The reason is simple: the function $p(t)$ is positive definite, while the function $q(t)$ does not have a definite sign. Hence, the growth of $p(t)$ for a contracting universe can be compensated by an increasing negative value of $q$, so that $R$  remains bounded. The well-known example of a contracting universe filled with a thermal radiation clearly illustrates this. The Kretschmann invariant for this solution
\be \n{KKK}
K=R_{\mu\nu\alpha\beta}R^{\mu\nu\alpha\beta}=12(q^2+p^2),
\ee
grows infinitely, so that the limiting curvature condition is violated.
In what follows we discuss constraints that can be used to prevent infinite curvature growth.

\section{Limiting curvature gravity}\label{Sec3}

\subsection{Action and gravity equations}

The limiting curvature gravity model is described by an action of the form
\be \n{action}
I=I_g+I_c+I_m ,
\ee
where $I_g$ is the Einstein-Hilbert action,
\be\n{Einstein}
I_g={1\over 2\kappa}\int \dd^4 x \sqrt{-g} R \hh \kappa=8\pi G/c^4,
\ee
and $I_m$ is the matter action. The term $I_c$ is the constraint action depending on the metric and the Lagrange multipliers that generate constraints on the curvature.
Later, we will specify the form of these constraints and the corresponding term of the action $I_c$. Now we just mention that the imposed restriction on the curvature has the form of inequalities (see discussion in Ref.\cite{Frolov:2021kcv}). They have the following properties. Before the curvature reaches its critical value the corresponding subcritical metric coincides with a standard solution of the unmodified Einstein equations. After the curvature reaches the critical value the solution becomes supercritical and it follows the modified constraint equations which prevent further growth of the curvature.

The variation of the total action (\ref{action}) over the metric gives the following ``gravity" equations:
\ba\label{dI}
\frac{2}{\sqrt{-g}}\frac{\delta I}{\delta g_{\mu\nu}}=0,
\ea
which have the form
\be\label{Einstein}
G_{\mu\nu}=\kappa(T_{\mu\nu}+X_{\mu\nu}).
\ee
Here $G_{\mu\nu}$ is the Einstein tensor and
\ba
&T^{\mu\nu}=\frac{2}{\sqrt{-g}}\frac{\delta I_m}{\delta g_{\mu\nu}}\hh
X^{\mu\nu}=\frac{2}{\sqrt{-g}}\frac{\delta I_c}{\delta g_{\mu\nu}},
\ea
are the stress-energy tensors of matter and  constraints. Besides the gravity equations the action $I$ also gives  additional equations for matter and constraints which are obtained by its variation over the Lagrange multipliers, that are variables additional to the metric. If these equations are satisfied and the actions $I_c$ and $I_m$ are covariant,
the following relations are valid:
\be\label{CT} 
T^{\mu\nu}{\!}_{;\nu}=0 \hh X^{\mu\nu}{\!}_{;\nu}=0 .
\ee
These conservation laws guarantee consistency of the gravitational equations.

\subsection{Reduced action and reduced gravity equations}

The tensor $\ts{G}$ for the metric (\ref{metr}) is
\bea\label{Gpq}
& -G^{\mu}_{\nu}=\mbox{diag}({\cal G},{\cal \hat{G}},{\cal \hat{G}},{\cal \hat{G}})\,  ,\\
& {\cal G}=3p\hh {\cal \hat{G}}=p+2q\, .
\eea
Similarly, the tensors $\ts{T}$ and $\ts{X}$, respecting the symmetry of the metric (\ref{metr}) have the form
\bea\label{Gpq}
& T^{\mu}_{\nu}=\mbox{diag}({\cal T},{\cal \hat{T}},{\cal \hat{T}},{\cal \hat{T}})\,  ,\\
& X^{\mu}_{\nu}=\mbox{diag}({\cal X},{\cal \hat{X}},{\cal \hat{X}},{\cal \hat{X}})\, .
\eea

Then the gravity equations (\ref{Einstein}) reduce to the following equations
\bea
&-{\cal G}=\kappa({\cal T}+{\cal X}) \, ,\n{EE1}\\
&-\hat{\cal G}=\kappa(\hat{\cal T}+\hat{\cal X}) \, .     \n{EE2}
\eea
Equation (\ref{CT}) and the conservation property of the Einstein tensor, $G^{\mu\nu}_{\ \ ;\nu}=0$, give
\bea\label{hatG}
& \hat{\cal{G}}={\cal G}+{1\over 3} {d{\cal G}\over d\ln a}\, ,\\
& \hat{\cal {X}}={\cal X}+{1\over 3} {d{\cal X}\over d\ln a}\, ,\n{hatX}\\
& \hat{\cal {T}}={\cal T}+{1\over 3} {d{\cal T}\over d\ln a}\, .\n{hatT}
\eea
In particular, these relations imply that if the temporal gravity equation (\ref{EE1}) is valid, the spatial gravity equation (\ref{EE2}) is also satisfied.

It is convenient to use symmetries of the cosmological spacetimes and  write down a reduced action for our gravitational system. Namely, it is easy to check that the 4D gravity equations \eq{dI} taken on the spacetime \eq{metr} can be equivalently derived from the dimensionally reduced action
\ba\n{III}
\mathbb{I}=2\pi^2\int \dd t\,a^3 b\, \mathbb{L}\, ,
\ea
where $\mathbb{L}$ is the Lagrangian of the system evaluated on the metric \eq{metr} and $2\pi^2$ is the volume of the  unit sphere $S^3$.

For example  the dimensionally reduced Einstein action is
\ba
\mathbb{I}_g=\frac{6\pi^2}{\kappa}\int \dd t\,a^3 b (p+q)\, .
\ea
The variation of this action over the temporal $b$  and spatial $a$ components of the metric after imposing the gauge-fixing condition $b=1$ gives
\ba
&\Big[ \frac{1}{2\pi a^3}\frac{\delta \mathbb{I}_g}{\delta b} \Big]_{b=1}
~=\frac{1}{\kappa}{\cal G}=\frac{3p}{\kappa}\, ,\\
&\Big[\frac{1}{6\pi b a^2}\frac{\delta \mathbb{I}_g}{\delta a}\Big]_{b=1}
=\frac{1}{\kappa}\hat{\cal G}=\frac{p+2q}{\kappa}\,.
\ea

\subsection{Thermal radiation}

We choose the stress-energy tensor of the matter in the form
\ba\nonumber
T^{\mu}_{\nu}&=\mbox{diag}(-\varepsilon, P,P,P) \hh {\cal T}=-\varepsilon \hh \hat{\cal T}=P,
\ea
where  $\varepsilon$ is the energy density and $P$ is the pressure.
In what follows we assume that the matter is hot thermal radiation with the equation of state $P={1\over 3}\varepsilon$.
The conservation law \eq{hatT} is satisfied if
\ba\label{Ca}
\varepsilon=C a^{-4} ,
\ea
where the factor $C$ is defined by the temperature $T$ of radiation and the number of massless degrees of freedom $n$.\footnote{Let us note that
the stress-energy tensor for thermal radiation can be derived from the reduced action
\ba
\mathbb{I}_m=v\int \dd t\,a^3 b\, \mathbb{L}_m \hh \mathbb{L}_m=-C a^{-4} .
\ea
}

For a closed homogeneous and isotopic universe filled with a thermal radiation the total energy $E$ and entropy $S$ of the universe are given by
\be\label{ES}
\varepsilon=\frac{E}{V}={\alpha_\ins{B}} T^4\hhh \frac{S}{V}=\frac{4}{3} {\alpha_\ins{B}} T^3\hhh {\alpha_\ins{B}}={n\pi^2 k_\ins{B}^4\over 30\hbar^3 c^3}\, .
\ee
Here $T$ is the radiation temperature and $V=2\pi^2 a^3$ is the volume of the closed universe. In the case of only electromagnetic radiation $n=2$. At high temperature many other fields become effectively massless. For example at the temperature corresponding to 300 GeV this number is about $n\approx 106$.
The relation \eq{ES} allows one to express the constant $C$ in Eq.(\ref{Ca}) in terms of the entropy $S$, which is a conserved quantity. One gets
\ba
(aT)^3=\frac{3S}{ 8\pi^2 {\alpha_\ins{B}}}
\ea
and, hence,
\ba\n{CCCC}
C=\nu \hbar c \Big( \frac{S}{k_\ins{B}}\Big)^{4/3}
\hh
\nu={3\over 16\pi^3}\Big( \frac{90}{n \pi}\Big)^{1/3}.
\ea
For pure electromagnetic radiation $n=2$ and we have $\nu\approx 0.014686 $.
Thus the value of the constant $C$ is defined by the entropy of the thermal gas in the Universe and can be estimated from observations \cite{Egan:2010}. For example the contribution of photons to the entropy is ${S/ k_\ins{B}}\sim 5.4\cdot10^{89}$ and, hence,  $C\sim 6.5\cdot 10^{117}\hbar c$, which is a huge number. The other massless particles like neutrinos contribute similar amounts to the entropy and energy density.

Let us note that at the stage of contraction the thermal radiation dominates. When the growing temperature becomes high enough particles with mass $m\ll T$ becomes ultrarelativistic and their contribution to the energy density is similar to the contribution of massless particles (photons), while the contribution of the particles with $m>T$ is relatively small. For this reason, in what follows we assume that the contracting universe is radiation dominated.

\section{Linear-in-curvature constraints}\label{Sec4}

\subsection{General form of linear constraints}

In order to control curvature growth  one can impose a restriction on the eigenvalues of  tensors constructed as a linear combination of the Ricci tensor $R^{\mu}_{\nu}$ and $R\delta^{\mu}_{\nu}$, where $R$ is a Ricci scalar. We call such constraints linear in curvature. Let us discuss the case of the linear constraints first and return to the discussion of other constraints constructed from curvature invariants later.

As earlier, we denote by $\ts{S}$ a traceless part of the Ricci tensor, and  $\ts{Q}={1\over 6} R \ts{I}$. Here $\ts{I}$ is a unit tensor. We denote
\be
\ts{Z}=c_S \ts{S}+ c_R \ts{Q}\, .
\ee
One has
\be
\ts{Z}=\mbox{diag}({\cal Z},\hat{\cal Z},\hat{\cal Z},\hat{\cal Z})\, .
\ee
The  eigenvalues of $\ts{S}$ and $\ts{Q}$ are linear functions of the quantities $p$ and $q$ defined by Eq.(\ref{defpq}).
Hence, $\ts{Z}$ has the same property.  Using Eqs.(\ref{Rpq}) and (\ref{Spq}), one gets
\bea\n{CONZ}
&&{\cal Z}=(c_R-{3\over 2}c_S)p+(c_R+{3\over 2}c_S)q\, ,\\
&&\hat{\cal Z}=(c_R+{1\over 2}c_S)p+(c_R-{1\over 2}c_S)q\, .
\eea
We shall restrict the curvature by imposing the conditions
\be\n{CZZ}
|{\cal Z}|\le {\Lambda} \hh  |\hat{\cal Z}|\le {\Lambda}\, .
\ee

\begin{figure}[!hbt]
    \centering
    \vspace{10pt}
    \includegraphics[width=0.25 \textwidth]{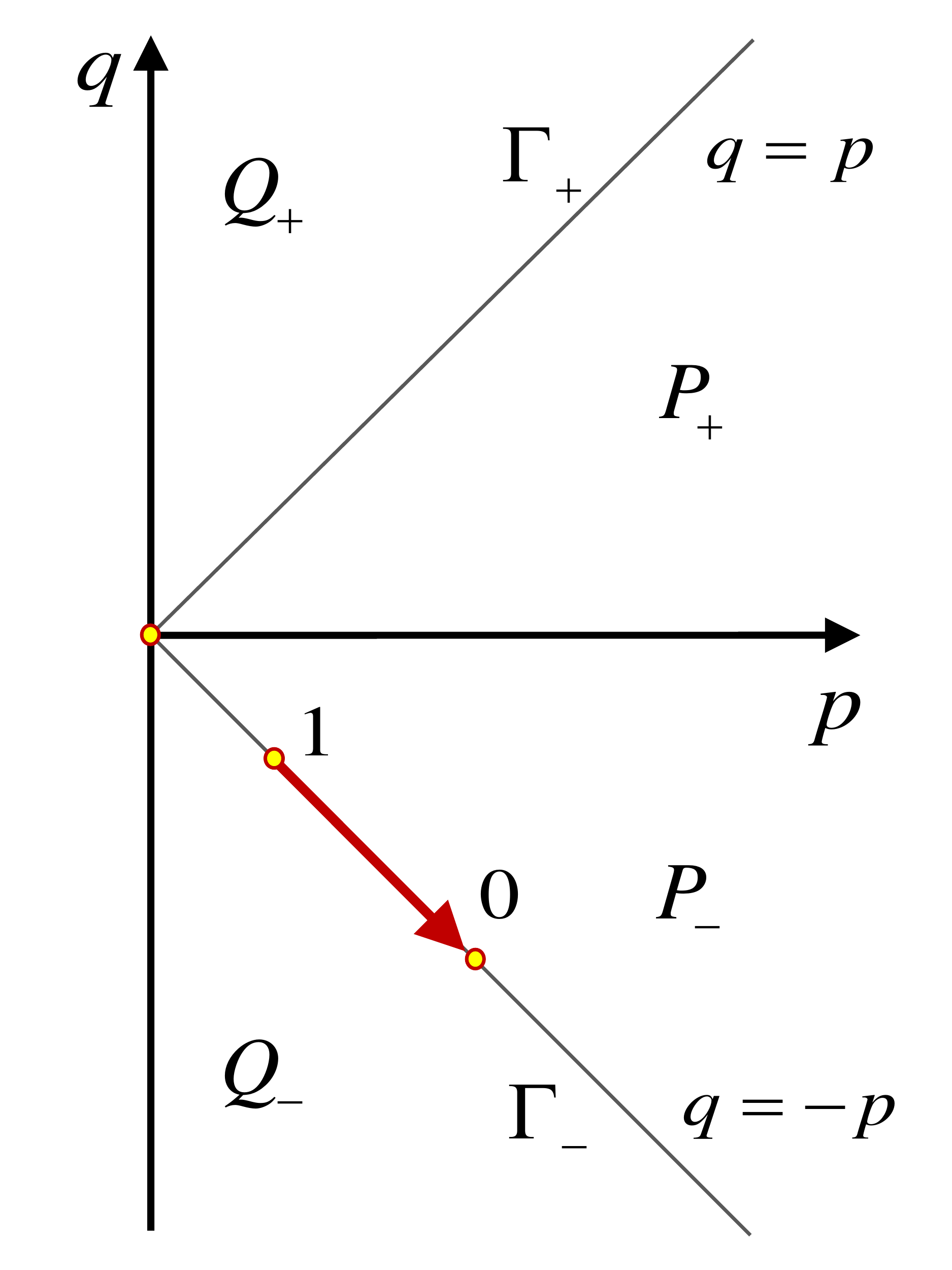}
    \caption{$p$-$q$ plane.}
    \label{Fig_pq1}
\end{figure}

Consider two-dimensional $(p,q)$ plane (see Fig.~\ref{Fig_pq1}).
A domain  in this  plane where the  inequalities (\ref{CZZ}) are satisfied is restricted by straight lines,
\be\n{ZZLL}
{\cal Z}=\pm  {\Lambda}\hh \hat{\cal Z}=\pm {\Lambda}\, .
\ee
We call of these lines as describing temporal and spatial constraints, respectively.

\subsection{Constraint action}

In order to provide the inequality constraints (\ref{CZZ}) we add the following expressions to the reduced action (\ref{III}):
\ba
&\mathbb{I}_c=2\pi^2\int \dd t\,a^3 b\, \mathbb{L}_c,\\
& \mathbb{L}_c=\mathbb{L}_0+\hat{\mathbb{L}}_+ +\hat{\mathbb{L}}_-\, ,\\
&{\mathbb{L}}_0=-{\chi} ({\cal Z}-{\Lambda}+ {\zeta}^2)\, ,\\
&\hat{\mathbb{L}}_{\pm}=-\hat{\chi}_{\pm} (\hat{\cal Z}\mp{\Lambda}\pm \hat{\zeta}_{\pm}^2)\, .
\ea
As we shall see later, the constraint ${\cal Z}=-  {\Lambda}$ does not give any restrictions on the physically interesting solutions. That is why we did not include the corresponding term in the action.

The variation of this action over the Lagrange multipliers gives the following equations
\bea
& {\cal Z}-{\Lambda}+ {\zeta}^2=0\hh {\chi}{\zeta}=0\, ,\\
&\hat{\cal Z}_\pm\mp {\Lambda}\mp \hat{\zeta}_{\pm}^2=0\hh \hat{\chi}_{\pm}\hat{\zeta}_{\pm}=0\, .
\eea
These equations imply that the system has two different regimes. In the subcritical regime where ${\chi}= \hat{\chi}_{\pm}=0$,   the nonvanishing parameters ${\zeta}$ and $\hat{\zeta}_{\pm}$ are defined in terms of ${\cal Z}$ and $\hat{\cal Z}_{\pm}$, respectively. In this regime the action $\mathbb{I}_c$ does not contribute to the gravity equations, so the evolution of the universe follows its standard solutions of the unmodified Einstein equations.

In the supercritical regime, when one of the constraint equations is saturated and the corresponding Lagrange multiplier ${\zeta}$ or $\hat{\zeta}_{\pm}$ becomes zero. This means that one of the constraint equations
\bea
& {\cal Z}-{\Lambda}=0\, ,\n{ZZLL1}\\
&\hat{\cal Z}_\pm\mp {\Lambda}=0 \n{ZZLL2}
\eea
is valid.
The corresponding control function $ {\chi}$ or $ \hat{\chi}_{\pm}$ becomes ``dynamical" and its  evolution in the supercritical regime is defined by the gravity equations.

The contributions ${\cal X}$ and $\hat{\cal X}$ of the constraint action $\mathbb{I}_c$ to the gravity equations  can be obtained as follows. Since  the constraint functions ${\cal Z}$ and $\hat{\cal Z}_\pm$ are linear combinations of the functions $p$ and $q$
given by Eq.(\ref{ppqq}), it is sufficient to find the variations over the metric of the following  reduced actions:
\bea
&\mathbb{I}_p=2\pi^2\int \dd t\,a^3 b\, u p \, ,\\
&\mathbb{I}_q=2\pi^2\int \dd t\,a^3 b\, u q \, , \\
&\mathbb{I}_u=2\pi^2\int \dd t\,a^3 b\, u  \, .
\eea
Here $u=u(t)$ stands for one of the control functions.
The variations of these reduced actions
over the temporal $b$  and spatial $a$ coefficients of the metric after imposing the gauge-fixing condition $b=1$ give
\ba\n{RED}
&{\cal X}_p\equiv \frac{1}{2\pi^2 a^3}\frac{\delta \mathbb{I}_p}{\delta b}=\Big(-\frac{\dot{a}^2}{a^2}+\frac{1}{a^2}\Big)u \, ,\\
&\hat{\cal X}_p\equiv\frac{1}{6\pi^2  b a^2}\frac{\delta \mathbb{I}_p}{\delta a}=-\frac{1}{3}\frac{\dot{a}}{a}\dot{u}+\frac{1}{6}\Big(-\frac{\dot{a}^2}{a^2}-2\frac{\ddot{a}}{a}+1\Big)u \, ,\\
&{\cal X}_q\equiv \frac{1}{2\pi^2  a^3}\frac{\delta \mathbb{I}_q}{\delta b}=\frac{\dot{a}}{a}\dot{u}+2\frac{\dot{a}^2}{a^2}u \, ,\\
&\hat{\cal X}_q\equiv\frac{1}{6\pi^2 b a^2}\frac{\delta \mathbb{I}_q}{\delta a}=\frac{1}{6}\ddot{u}+\frac{2}{3}\frac{\dot{a}}{a}\dot{u}+\frac{1}{3}\Big(\frac{\dot{a}^2}{a^2}+\frac{\ddot{a}}{a}\Big)u \, ,\\
&{\cal X}_u\equiv \frac{1}{2\pi^2  a^3}\frac{\delta \mathbb{I}_u}{\delta b}=u \, ,\\
&\hat{\cal X}_u\equiv\frac{1}{6\pi^2 b a^2}\frac{\delta \mathbb{I}_u}{\delta a}=\frac{1}{2}u \, .
\ea

\section{Subcritical solutions}\label{Sec5}

At this stage the control functions $ {\chi}$ or $ \hat{\chi}_{\pm}$ vanish and the functions  ${\zeta}$ and $\hat{\zeta}_{\pm}$ drop out of the equations. As a result the standard Einstein equations govern the dynamics of the radiation-dominated universe. The temporal Einstein equation is
\ba\label{tildeGT}
{\cal G}=-\kappa {\cal T}\, .
\ea
In an explicit form this equation reads
\ba\label{subcritical0}
&\frac{\dot{a}^2+1}{a^2}=\frac{\kappa C}{3 a^4}\, .
\ea
Here we fixed the gauge and put $b=1$. In this gauge we have
\be \n{defpq}
p=\frac{\dot{a}^2+1}{a^2}\hh q=\frac{\ddot{a}}{a}\, .
\ee
The spatial Einstein equation is the consequence of Eq.(\ref{subcritical0}) and it reduces to
\ba\label{subcritical1}
&{\ddot{a}\over a}=-{\kappa C\over 3 a^4} .
\ea
Note that the scalar curvature during this stage of evolution vanishes.
This means that a point representing the state of the universe in the $(p,q)$ plane moves along the line $\Gamma_-$ where $q=-p$ (see Fig.~\ref{Fig_pq1}).

A  solution of the equation Eq.\eq{subcritical0} is well known (see, e.g., Ref.\cite{Landau:1975pou}). It has the form
\be\label{at}
a=\sqrt{-2t a_\ins{max}-t^2} \hh a_\ins{max}=\sqrt{\frac{\kappa C}{3}} ,
\ee
where the integration constant is fixed by the  condition $a(t=0)=0$.  Here $a_\ins{max}$ is the maximal value of the scale factor of the Friedmann universe. Note that $t<0$ during the contraction stage.

During the collapse of the universe its scale parameter $a(t)$ decreases and the energy density of matter grows.
This subcritical regime continues until the curvature reaches the critical value at which the supercritical regime starts. It happens when one of the constraint equations (\ref{ZZLL1})--(\ref{ZZLL2}) is satisfied. We denote the corresponding critical value of $p$ by $\lm$. Thus, the supercritical solution starts at the point $(p=\lm,q=-\lm)$.
Let us find the parameters $a_0$ and $\dot{a}_0$ of the contracting universe at this point.
For $p=\lm$ Eqs.(\ref{subcritical0}) and  (\ref{defpq}) give
\be
{a}_0=\left({ \kappa C\over 3\lm}\right)^{1/4}\, .
\ee
Using the definition of $p$ one gets
\be
 \dot{a}_0=-\left[a_\ins{max}\lm^{\frac{1}{2}}-1\right]^{\frac{1}{2}}\, .
\ee
In the latter relation  we choose a minus sign since we assume that the universe is initially contracting. The moment of transition to the supercritical stage is
\ba
t_{0}=-\frac{a_\ins{max}}{\lm^{\frac{1}{2}}}\left[\left(1-\frac{1}{a_\ins{max}\lm^{\frac{1}{2}}}\right)^{\frac{1}{2}}-1\right].
\ea


\section{Linear-in-curvature constraints: Supercritical solutions}\label{Sec6}

\subsection{General remarks}

We are looking for  constraints that restrict the curvature, so that during all of the subsequent evolution of the universe after it enters a supercritical regime the curvature remains finite and restricted by a chosen universal value. To characterize the value of the curvature one can use, for example, the Kretschmann invariant $K$ Eq.(\ref{KKK}).

Let $a(t)$ be a solution for a scale function which determines the size of the universe.
For a general constraint after the solution enters the supercritical regime
it may terminate at some finite time $t_s$. This may happen if the differential equation for $a(t)$ determined by the constraint has a singular point which prevents an extension of the solution beyond time $t_s$. We call such a constraint, which does not allow a complete description of the evolution of the universe, a singular one. In what follows we shall not consider such constraints. Namely, we assume the following.
\begin{itemize}
\item The supercritical solution is not terminated at finite time $t$.
\item The constraint guarantees that during the supercritical regime the Kretschmann invariant is uniformly restricted by some universal value which does not depend on the  parameters of the  solution.
\end{itemize}
We call such a  constraint a regular one. For this type of constraint, the corresponding supercritical solution can either be continued to $t\to\infty$ or slip back to its subcritical phase. In principal, if there exist several constraints, the supercritical solution can also slip between them.

 We  assume that the constraint line $C$ intersects $\Gamma_-$ at  $p=\lm$ where the solution enters the supercritical regime. For the contracting universe $\dot{a}<0$ at this point.
Using the definition (\ref{defpq}) for $p$ and $q$, one can obtain the following equation:
\be\n{dpt}
\frac{d p}{dt}=2{\dot{a}\over a}(q-p)\, .
\ee
While a point on $C$ representing a supercritical solution is located in the domain where $q<0$ the negative value of $\dot{a}$ can only increase. Thus in this domain $\dot{a}<0$. One also has $q-p<0$ in the domain of the $(p,q)$ plane below $\Gamma_-$. Equation (\ref{dpt}) shows that under these conditions the point representing the supercritical contracting universe in the $(p,q)$ plane can move only with an increase of the parameter $p$.

\subsection{Regular linear constraints}

\begin{figure}[!hbt]
    \centering
    \vspace{10pt}
    \includegraphics[width=0.25 \textwidth]{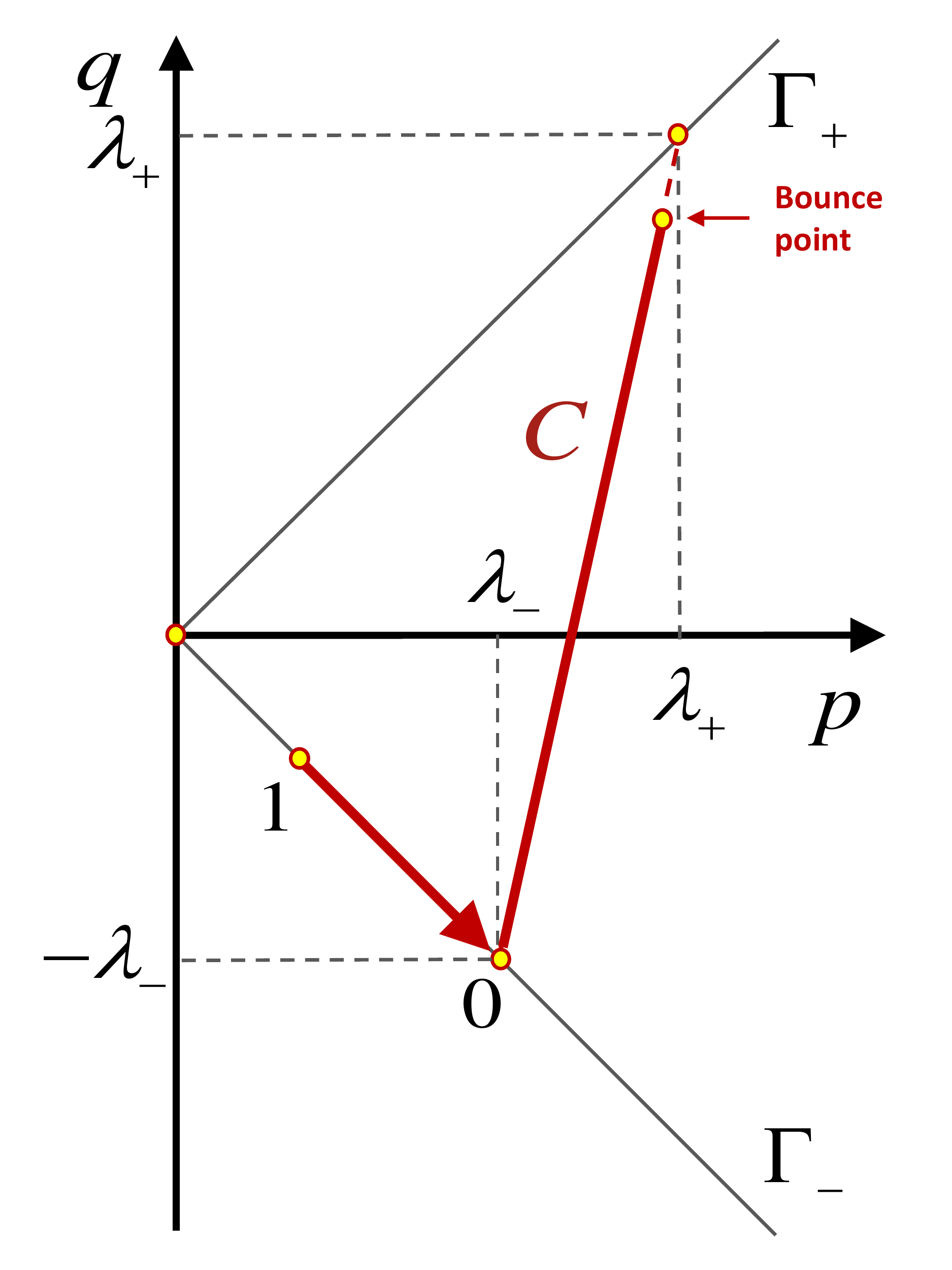}
    \caption{Linear constraint in the $p$-$q$ plane.}
    \label{pq_0}
\end{figure}

To describe the evolution of the universe we use the two-dimensional $(p,q)$ plane.
Let $\Gamma_{\pm}$ be two lines on this plane defined by the equations $q=\pm p$, respectively.
The subcritical evolution of a contracting radiation-dominated universe  is represented by the interval on $\Gamma_-$ (see Fig.~\ref{pq_0}).

Let $C$ be a straight  line representing the linear-in-curvature constraint. We assume that this line intersects $\Gamma_-$ at a point $0$ and write its equation in the form
\be \n{linpq}
p-\mu q={\Lambda}\, .
\ee
The parameter ${\Lambda}$ has dimensions of [length]$^{-2}$ and characterizes the value of the limiting curvature. Since the left-hand side of Eq.(\ref{linpq}) is positive at the point $0$, ${\Lambda}$ is chosen to be positive as well.
In the presence of the constraint  (\ref{linpq}) a point representing the evolution of the universe after it reaches the point 0 starts its motion along a constraint  $C$. Let us discuss the corresponding supercritical solution.

\subsubsection{Negative-$\mu$ case}

Let us first assume that the parameter $\mu$ in Eq.(\ref{linpq}) is negative.
Then, $dp/dq=\mu<0$.  If $C$ does not meet another constraint,  both $p$ and $|q|$ along $C$ grow infinitely and as a result the Kretschmann invariant grows as well. This means that such a constraint is not regular. For this reason we assume that $\mu>0$.

\subsubsection{$\mu>1$ case}

Let us consider the case where $\mu>1$. At the point $0$ where the supercritical regime starts  $p-\mu q>0$ and
one has
\be
p_-=-q_-=\lm\equiv {{\Lambda}\over 1+\mu}\, .
\ee
After this, a representative point which is moving with the increase of $p$ enters the domain above $\Gamma_-$ and remains there since the corresponding line $C$ cannot intersect $\Gamma_+$.

To find how the scale factor behaves in this case we rewrite Eq.(\ref{dpt}) in the form
\be
\frac{d p}{d\ln(a^2/a_0^2)}=q-p\, .
\ee
Here $a_0$ is the size of the universe at the beginning of the supercritical regime when $p=\lm$.
Integrating this equation with the imposed initial conditions we get
\be\n{App}
a=a_0 \exp(-F)\hh
F=\frac{1}{2}\int_{\lm}^p \frac{dp}{p-q(p)}\, .
\ee
Here $F$ is the expansion factor and the function $q(p)$ is defined by the Eq.(\ref{linpq}).
The integral can be easily calculated and one has
\be
F=\frac{\mu}{2(\mu-1)}\ln\left[ {(\mu-1)p+(1+\mu)\lm\over 2\mu \lm}\right]\, .
\ee
The integration constant is chosen so that $F|_{p=\lm}=0$.
Thus the relation between $a$ and $p$ takes the form
\be\n{A0}
a=a_0\left[{2\mu \lm\over (\mu-1)p+(1+\mu)\lm}\right]^{ \mu\over 2(\mu-1)}\, .
\ee
Since $\mu>1$ and $p$ grows monotonically the scale function $a$
monotonically decreases. The Kretschmann invariant grows infinitely along the constraint while the size of the universe shrinks. Thus such a constraint is not regular.

\subsubsection{Case $0<\mu<1$.}

Let us consider the last case where $0<\mu<1$. In this case the constraint line $C$ crosses $\Gamma_+$. At the point of the intersection $p_+=q_+\equiv \lp$
\be
\lp={{\Lambda}\over 1-\mu}>\lm  \, .
\ee
One also has
\be\label{mu0}
\mu={\lp-\lm\over \lp+\lm}\, .
\ee
Let us introduce the dimensionless quantities
\be
\bar{p}=\displaystyle\frac{p}{\lp}\hhh \bar{q}=\frac{q}{\lp}\hhh \alpha=\sqrt{\lp}\, a\hhh
\tau=\sqrt{\lp}\, t\, .
\ee
Then one has
\be \n{zpq}
\bar{p}-1=\mu(\bar{q}-1)\hh \bar{q}-\bar{p}={1-\mu\over \mu}(\bar{p}-1)\, .
\ee

Equation (\ref{App}) can be used to find a relation between $\bp$ and $\alpha$. It is sufficient to substitute into the expression for $F$ the function $q(p)$ defined by Eq.(\ref{zpq}). The integral can be easily calculated and one has
\be
F=-\frac{\mu}{2(1-\mu)}\ln\left[{ (1+\mu) \over 2\mu}(1-\bar{p})\right]\, .
\ee
Thus the relation between $\alpha$ and $\bar{p}$ takes the form
\be\n{aa0}
\alpha=\alpha_0\exp(-F)=\alpha_0
\left[{ (1+\mu) \over 2\mu}(1-\bar{p})\right]^{\mu \over 2(1-\mu)}\, .
\ee
By inverting this relation we find $\bar{p}$ as a function of $\alpha$, and then by using Eq.\eq{zpq} we also compute $\bar{q}$ as a function of $\alpha$,
\ba\label{pmu}
\bar{p}=1-\frac{2\mu}{1+\mu}\Big(\frac{\alpha}{\alpha_0}\Big)^{\frac{2(1-\mu)}{\mu}} ,
\ea
\ba\label{qmu}
\bar{q}=1-\frac{2}{1+\mu}\Big(\frac{\alpha}{\alpha_0}\Big)^{\frac{2(1-\mu)}{\mu}} .
\ea

Let us demonstrate that for a chosen linear constraint a contracting universe always has a bounce. Let us assume that such a bounce exists. At this point, where  $\dot{a}=0$, the universe has a minimal size, which we denote by  $\alpha_b$.
Let  $\bar{p}_b$ be the corresponding value of the parameter $\bar{p}$ at this point.  Using Eq.(\ref{defpq}), one has
\be\label{palpha}
\bar{p}^{1/2}_b \alpha_b=1 .
\ee
After using Eq.(\ref{aa0}), this condition takes the form
\be\n{ZZ}
\frac{1}{\bar{p}_b} \left[{2\mu\over (1+\mu)(1-\bar{p}_b)}\right]^{\mu \over (1-\mu)}=\alpha_0^2\, .
\ee
For every $0<\mu<1$ the function on the left-hand side of this relation grows infinitely when $\bar{p}_b\to 1$.
This means that for an arbitrarily large $\alpha_0$ Eq.(\ref{ZZ}) has a solution. In other words the universe has a bounce. For large $\alpha_0$, this happens when $\bar{p}_b$ is close to $1$.
In this case one can omit the term $\bar{p}^{1/2}_b $ in Eq.\eq{palpha}.

Equation (\ref{ZZ}) allows one to express $\bp$ as a function of $\alpha$. After substituting  this expression
into the relation
\be\label{alphaprime}
(\alpha')^2=\alpha^2 \bp-1\, ,
\ee
one obtains the equation that determines the evolution of the universe in the supercritical regime.
Here a prime is a derivative with respect to $\tau$.

After the size of the universe reaches the minimal value $a_b$ it expands again.  The point representing it in the $(p,q)$ plane moves again along the line $C$ but now in the opposite direction with the decreasing value of $\bp$. At the point where  the solution intersects $\Gamma_-$ it can leave the supercritical phase. Such a solution describes an expanding  universe filled with thermal radiation. Let us emphasize that during its evolution in the supercritical regime the value of the Kretschmann invariant remains uniformly restricted. Thus the linear constraint (\ref{linpq}) with $0<\mu<1$ is regular.

\subsection{Temporal and spatial constraints}

Both temporal and spatial curvature constraints can be written in a form similar to Eq.(\ref{linpq}).
Let us first apply the results of the previous section to the temporal constraint.
In order to present it in the form (\ref{linpq}) it is sufficient to choose the coefficients $c_R$ and $c_S$ in Eq.(\ref{CONZ}) in the form
\be
c_R={1\over 2}(1-\mu)\hh c_S=-{1\over 2}(1+\mu)\, .
\ee
Then one has
\bea \label{TC}
 &&Z={\Lambda}\hh Z\equiv p-\mu q \,  ,\\
&&\hat{Z}=\pm {\Lambda}\hh \hat{Z}\equiv {1\over 3}  \left[ (1-2\mu)p+(2-\mu)q \right]\, . \label{SC}
\eea
We choose $0<\mu<1$. Then the temporal constraint (\ref{TC}) intersects $\Gamma_-$ at the point $-q=p=\lm$, where $\lm={\Lambda}/(1+\mu)$. The  evolution of the universe is represented in the $(p,q)$ plane by two intervals: one is the interval along $\Gamma_-$ until the point 0 where $\bp=\lm$, and the other is the interval on the constraint line $C$ from 0 until the turning point $(\lp(1-\Delta p),\lp(1-\mu^{-1}\Delta p)$. For a large initial size of the universe $\alpha_0\gg 1$ the positive quantity $\Delta p$ is small. After the turning point the universe moves back along $C$  up to the point $0$, where it can slip to the subcritical solution describing an expanding universe.

Let us show that for this motion the spatial constraints (\ref{SC}) are always satisfied. The spatial constraints define a domain in the $(p$-$q)$ plane, where the corresponding functions of curvatures are restricted. This domain is a strip located between the straight lines  $\hat{Z}= {\Lambda}$ and $\hat{Z}=- {\Lambda}$. We call these lines the upper and lower bounds, respectively.
We denote by $\hat{p}_{\pm}$ the coordinates $p$ of the points where the
spatial constraint $\hat{Z}$ intersects $\Gamma_{\pm}$ lines. At these points one has
\bea
&&\hat{Z}_+ \equiv \hat{Z}(\hat{p}_+)=(1-\mu)\hat{p}_+\, ,\\
&&\hat{Z}_- \equiv \hat{Z}(\hat{p}_-)=-{1\over 3}(1+\mu)\hat{p}_-\, .
\eea
One can check that
\be
\hat{p}_+=\pm \lp\hh \hat{p}_-=\mp 3 \lm\, .
\ee
In these relations the upper signs stand for the upper bound constraint and the lower signs stand for the lower bound constraint. It is easy to check that the curve representing the evolution of the universe obeying the temporal constraint always lies inside the domain restricted by upper and lower bound lines. In other words, the spatial constraints do not impose any restriction on the evolution of the universe and hence can be ignored.

\subsection{Evolution of the control function $\chi$}

Let us now discuss the gravity equations (\ref{EE1})--(\ref{EE2}). As we already mentioned, as a result of the conservation law the second of these equations (the spatial equation) is satisfied if the first (temporal) equation is valid. We rewrite the latter in the form
\be \n{contr}
{\cal X}=-({1\over \kappa}{\cal G}+{\cal T})\emph{}\, .
\ee
Using expressions for ${\cal G}$ and ${\cal T}$, one gets
\be\n{XXX}
-({1\over \kappa}{\cal G}+{\cal T})={C\over a^4}-{3\over \kappa}{\dot{a}^2+1\over a^2}\, .
\ee

The control function $\chi$ vanishes in the subcritical regime where ${\cal X}$ is also zero. Equation (\ref{contr}) determines the evolution of the control function in the supercritical regime. In such a case one can put $\zeta=0$ and use the reduced action
\ba
\mathbb{I}_c=2\pi^2\int \dd t\,a^3 b\, \mathbb{L}_0      \hh
{\mathbb{L}}_0=-{\chi} (p-\mu q-{\Lambda}) .
\ea
Taking the variation of the reduced action $\mathbb{I}_c$ over $b$ and putting $b=1$, one gets
\be
{\cal X}=-{\cal X}_p+\mu {\cal X}_q+{\Lambda}{\cal X}_{\chi}\, .
\ee
Using Eq.(\ref{RED}), one obtains
\be\n{YYY}
{\cal X}=\mu {\dot{a}\over a}\dot{\chi}+\left[(1+\mu){\dot{a}^2\over a^2}-{1\over a^2}+{\Lambda}\right]\chi\, .
\ee
Combining Eqs.(\ref{contr}), (\ref{XXX}), and (\ref{YYY}), one can write the equation for the control function $\chi$ in the following dimensionless form:
\be
\mu {\alpha'\over \alpha}\bc'+\left[ (1+\mu){\alpha'^2\over \alpha^2}-{1\over \alpha^2}+1-\mu\right]\bc=
{1-\mu\over 1+\mu}{\alpha_0^4\over\alpha^4}-\bp\, .
\ee
Here $\bar{\chi}=\chi/\lp$.
This  equation determines the time dependence of the control function $\bar{\chi}$ in the supercritical regime. For a given $\alpha(\tau)$ this is a first-order linear inhomogeneous ordinary differential equation (ODE).
This equation can be written in such a form that the control function $\bar{\chi}$ explicitly depends only on  $\alpha$,
\bea\label{eqchiB}
\mu\Big(\bar{p}-\frac{1}{\alpha^2}\Big)\frac{d\bar{\chi}}{d\ln \alpha}
&+\Big[(\bar{p}+1)(1-\mu)-(2-\mu)\frac{1}{\alpha^2}\Big]\bar{\chi}\nonumber\\
&=\frac{1-\mu}{1+\mu}\frac{\alpha_0^4}{\alpha^4}-\bar{p} \, ,\label{eqchiB}
\eea
where $\bp(\alpha)$ and $\bq(\alpha)$ are given by Eqs.\eq{pmu}--\eq{qmu}.
Therefore, the time dependence of the control function is uniquely determined by the time dependence of the scale parameter $\alpha(\tau)$. The evolution of the metric is symmetric with respect to the time reflection $(\tau-\tau_\ins{b})\to -(\tau-\tau_\ins{b})$ at bounce time $\tau_\ins{b}$. It is shown in the Appendix that there exists a solution for $\bar{\chi}$ which has the same property.\footnote{Let us note that a similar property is valid not only for linear in curvature constraints but also for a wider class of nonlinear constraints  (see the Appendix).}

\subsection{Phase diagram}

In the previous discussion we focused on the description of the evolution of the universe by using $(p,q)$ planes. Let us now describe this evolution by using the phase-space diagrams. Let us consider a two-dimensional space with coordinates $(\alpha,\alpha')$.
Equation (\ref{zpq}) can be written in the form
\be \n{EQQ}
\alpha''={1\over \mu}\left[ {\alpha'^2+1\over \alpha}-(1-\mu)\alpha\right]\, .
\ee
This second-order ODE is equivalent to the following system of two first-order equations:
\bea  \n{sys1}
&&\alpha'=z\, ,\\
&&z'={1\over \mu}\left[ {z^2+1\over \alpha}-(1-\mu)\alpha\right]\, .  \n{sys2}
\eea

\begin{figure}[!hbt]
    \centering
    \vspace{10pt}
    \includegraphics[width=0.50 \textwidth]{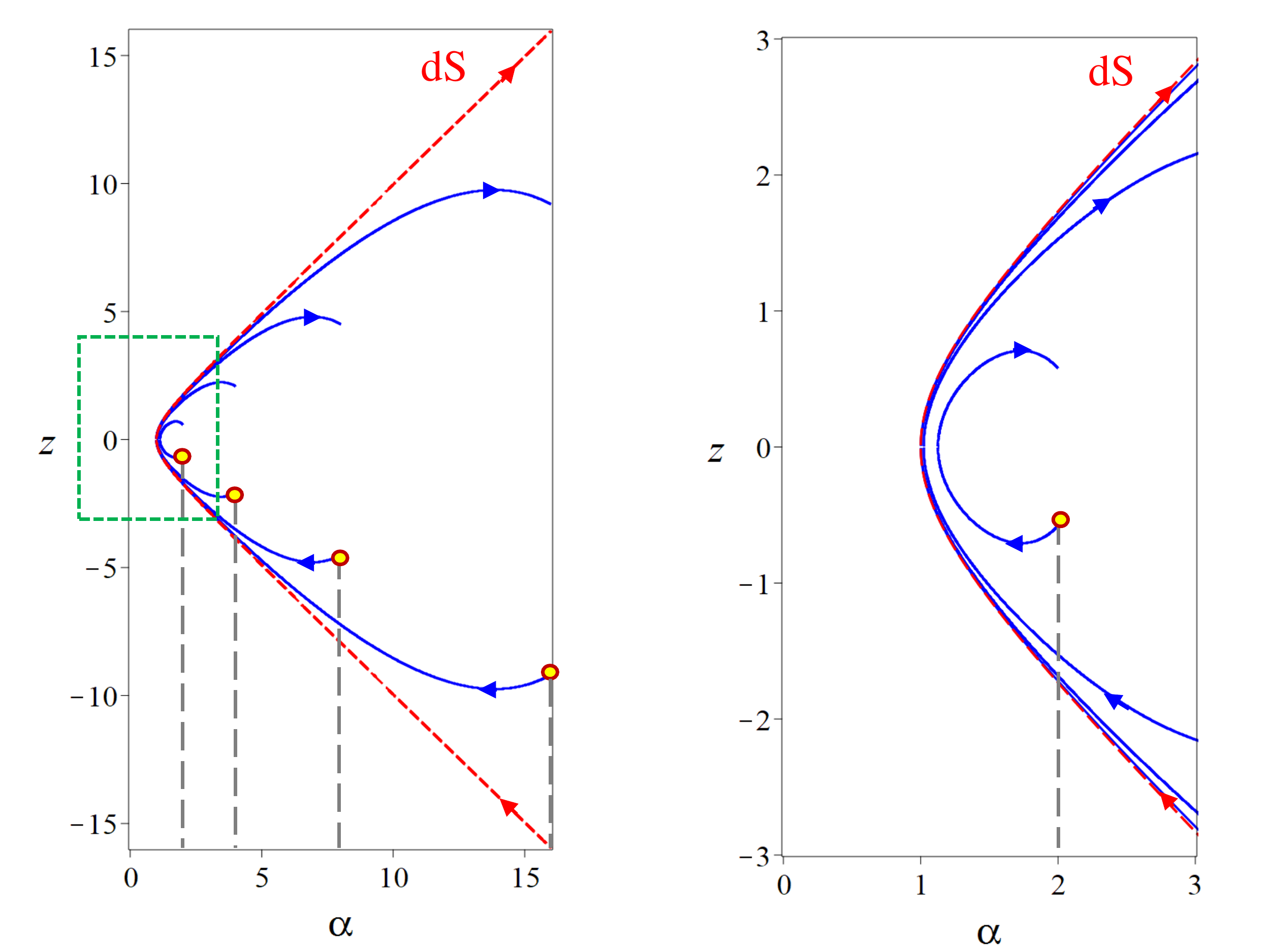}
    \caption{Phase plane $(\alpha,z)$ for $\mu=0.5$. For illustration purpose only we put rather moderate values for the initial values $\alpha_0=2,4,8,16$.
    At much larger $\alpha_0$ values the trajectories asymptotically approach the de Sitter hyperbola. The right panel depicts trajectories in the vicinity of the bounce point.}
    \label{Alpha_1}
\end{figure}

Phase diagrams for the system (\ref{sys1})--(\ref{sys2}) are shown in Fig.~\ref{Alpha_1}. A dashed line represents the de Sitter solution which approximates a general solution near the turning points.

\begin{figure}[!hbt]
    \centering
    \vspace{10pt}
    \includegraphics[width=0.45 \textwidth]{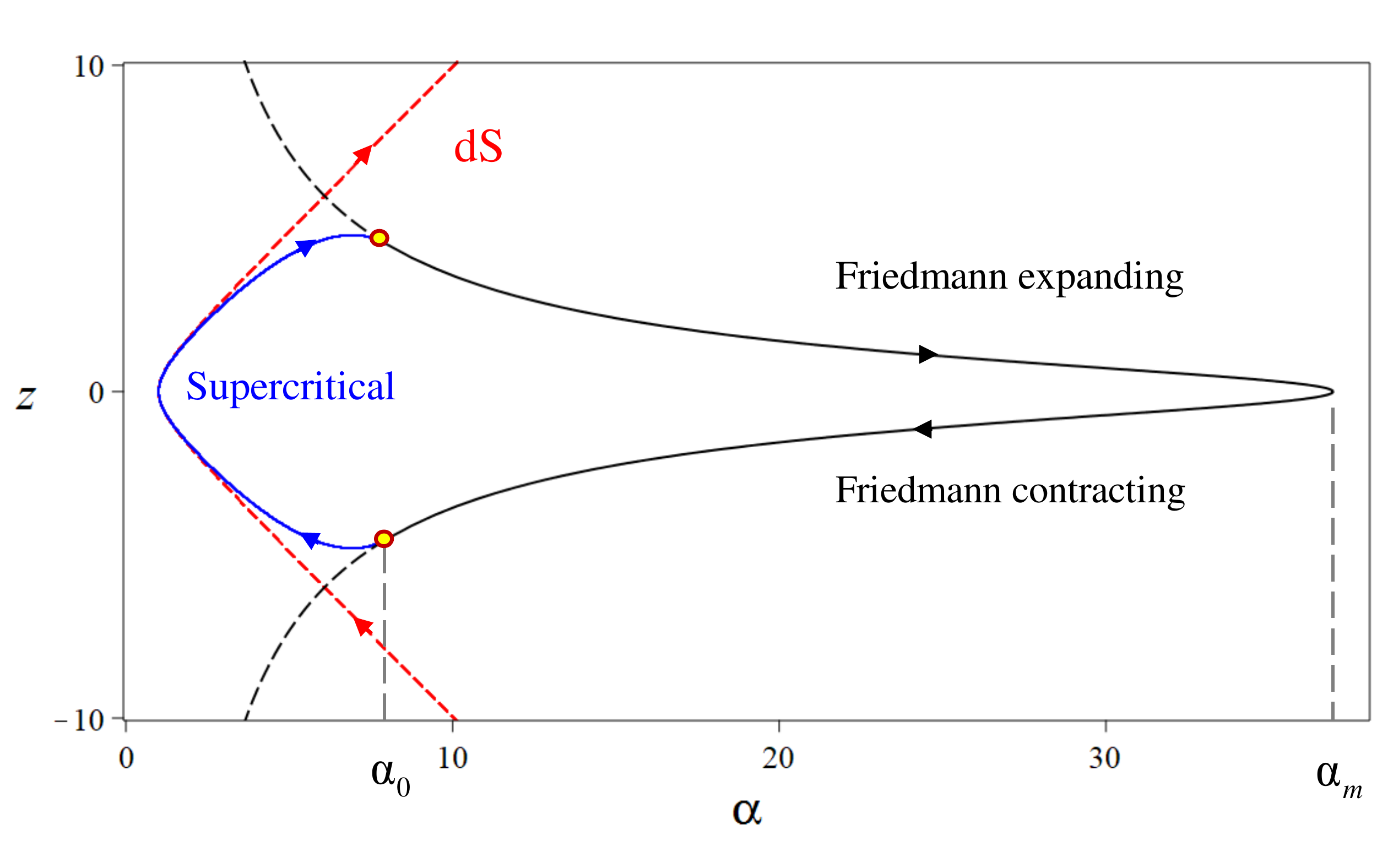}
    \caption{Illustration of a typical phase diagram for the whole trajectory which besides the supercritical stage contains subcritical Friedmann contracting and expanding stages. We use the parameters $\mu=0.5$ and $\alpha_0=8$. The blue curve depicts the supercritical stage of evolution $\alpha_0\to\alpha_\ins{b}\to\alpha_0$. Black lines represent the trajectories of contracting and expanding the radiation-dominated stages. Their dashed parts depict how the universe would evolve without the limiting curvature constraint.}
    \label{Alpha_1F}
\end{figure}

The dynamics of the universe is described by the system (\ref{sys1})--(\ref{sys2}) with the initial condition
\bea
&\alpha(\tau_0)=\alpha_0
=\Big[\frac{(1+\mu)\bar{\kappa}C}{(1-\mu)3}\Big]^{\frac{1}{4}} \, ,\\
& z(\tau_0)=\sqrt{ \frac{1-\mu}{1+\mu}\alpha_0^2-1}\, .
\eea

Let us denote
\be
{\Lambda}={1\over \ell^2}\, .
\ee
The parameter $\ell$ which has dimensions of length is the critical length corresponding to the limiting curvature ${\Lambda}$. Then, by using Eq.(\ref{CCCC}) one can write $\alpha_0$ in the form
\ba
&&\alpha_0=\beta \sqrt{l_\ins{Pl}\over \ell} \left( {S\over k\ins{B}}\right)^{1/3}\, ,\\
&&\beta=\left[\frac{1+\mu}{2\pi^2(1-\mu)^2}\right]^{1/4} \left(\frac{90}{n\pi}\right)^{1/12}\, .
\ea
Here $l_\ins{Pl}$ is the Planck length.
An effective curvature radius during inflation that is consistent with observations is usually considered to be in the range $10^5$--$10^9 l_\ins{Pl}$.
If one chooses the critical length $\ell$ to be of the same order of magnitude, then $(l_\ins{Pl}/\ell)^{1/2}\approx 3\cdot10^{-3}$--$3\cdot10^{-5}$. Since the entropy of our Universe  is large ($S/k\approx 10^{90}$),  the parameter $\alpha_0$ is also very large $\alpha_0\approx 10^{25}$--$10^{27}$.

The minimal value of dimensionless radius is achieved at the bounce point $\alpha_\ins{b}$. For every choice of parameters of the system  $\mu$ and $\alpha_0$ the bounce point $\alpha_\ins{b}$ can be found from the  equation
\ba
\frac{1}{\alpha_\ins{b}^2}=1-\frac{2\mu}{1+\mu}\Big(\frac{\alpha_\ins{b}}{\alpha_0}\Big)^{\frac{2(1-\mu)}{\mu}}
\ea
Because $\alpha_0$ is assumed to be very large, the condition
\ba
\alpha_0\gg e^{\frac{\mu}{2(1-\mu)}},
\ea
is satisfied for all $0<\mu<\mu_\ins{max}$, where $\mu_\ins{max}$ is close to 1.  In this case the bounce happens very close to unity
\ba
\alpha_\ins{b}\approx 1+\frac{\mu}{1+\mu}\alpha_0^{-\frac{2(1-\mu)}{\mu}} .
\ea
For example for $\mu=0.97$ one has $\alpha_\ins{b}\approx 1.01$. For smaller values of $\mu$ the bounce radius becomes exponentially close to 1.
In the range of $\alpha_0\approx 10^{25}$--$10^{27}$ the corresponding number of $e$-folds $N=\ln(\alpha_0/\alpha_\ins{b})$ is about
$
57<N<62 .
$
Recall that during the supercritical stage the universe first contracts from $a_0$ to $a_\ins{b}$. Then, the inflationary stage begins and it expands back to $a_0$ with the $e$-fold number $N$. After that, the Friedmann big bang expansion governed by the standard Einstein equations continues, as depicted in Fig.\,\ref{Alpha_1F}.

In the vicinity of the bounce point the trajectory of the supercritical evolution is very close to the de Sitter spacetime (see Fig.~\ref{Alpha_1}). For very large values of $\alpha_0$, the supercritical trajectory spends most of its time close to the de Sitter solution.  Qualitatively the de Sitter--like behavior happens when the acceleration $\ddot{a}$ changes sign from negative to positive. This is because the effective positive cosmological constant corresponds to repulsive gravity effects. Thus, the criterion of closeness of a supercritical solution to the de Sitter metric is that $q>0$.
The scale factor $a_\ins{dS}$ when $q=0$ can be estimated as follows. At this point $\bar{p}=1-\mu$, and using Eq.(\ref{aa0}) one gets
\ba
\frac{a_\ins{dS}}{a_0}=\frac{\alpha_\ins{dS}}{\alpha_0}=\Big(\frac{1+\mu}{2}\Big)^{\frac{\mu}{2(1-\mu)}} .
\ea
For all values of $0\le \mu\le 1$ we have $0.778 < a_\ins{dS}/a_0 \le1$, i.e., the de Sitter--like stage always happens very soon after the beginning of the supercritical regime.

\subsection{Effective Lagrangian}

\begin{figure}[!hbt]
    \centering
    \vspace{10pt}
    \includegraphics[width=0.3 \textwidth]{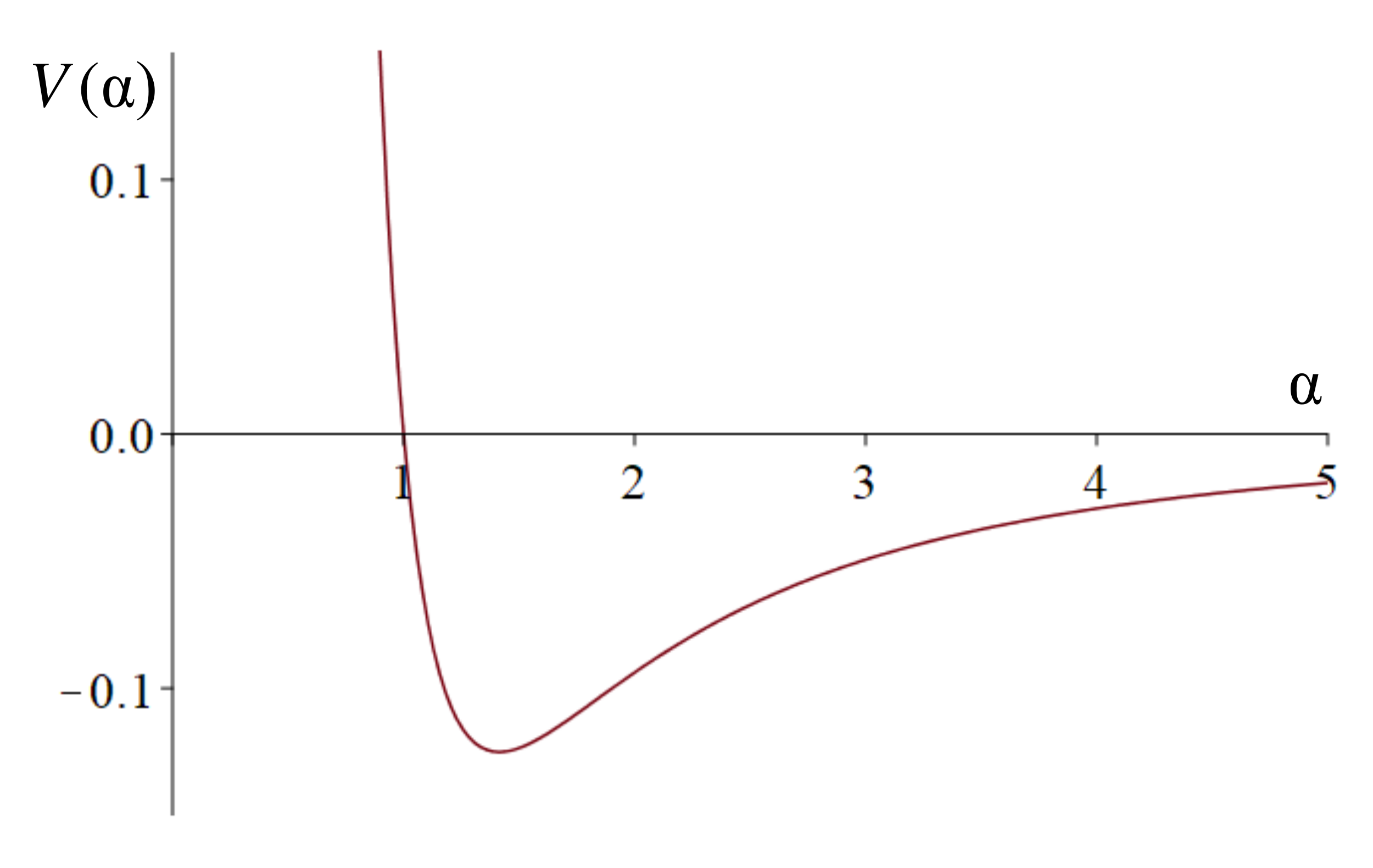}
    \caption{Potential $V(\alpha)$ for $\mu=0.5$.}
    \label{VVV}
\end{figure}

Let us note that the Eq.(\ref{EQQ}) coincides with the Euler-Lagrange equation for the following Lagrangian
\bea
&&L={1\over 2}m(\alpha)\alpha'^2-V(\alpha)\, ,\n{LAGR}\\
&&m(\alpha)=\alpha^{-2/\mu}\, ,\\
&& V(\alpha)={1\over 2}\alpha^{-2/\mu}(1-\alpha^2)\, .
\eea
Figure~\ref{VVV} shows the potential $V$ as a function of its argument $\alpha$.

This Lagrangian determines the dynamics of the dimensionless scale factor $\alpha$ during the supercritical regime.
The initial conditions for such motion are
\bea
&\alpha=\alpha_0\, ,\\
&\alpha'=\alpha'_0\equiv \sqrt{ \frac{1-\mu}{1+\mu}\alpha_0^2-1}\, .
\eea
Since the Lagrangian (\ref{LAGR}) does not contain an explicit dependence on time $\tau$, the ``energy"
\be
E=\alpha' {\partial L\over \partial \alpha'}-L={1\over 2}m(\alpha)\alpha'^2+V(\alpha)\,
\ee
is conserved. Using the initial conditions one can find
\be
E=-\frac{\mu}{1+\mu}\alpha_0^{-\frac{2(1-\mu)}{\mu}}\, .
\ee
The motion with negative energy $E$ in the potential $V(\alpha)$ is bound. In particular, $\alpha$ always has a ``left" turning point where it takes the minimal value $\alpha_b$. This conclusion is in agreement with the above  general analysis of the evolution of the scale factor in the theory of limiting curvature with linear-in-curvature constraints.
Let us notice that the solution also has a ``right" turning point where the scale factor reaches its maximal value,
\be
\alpha_*=\alpha_0 \left({ 1+\mu\over 2\mu}\right)^{\mu \over 2(1-\mu)}\, .
\ee
If the coefficient $\mu$ is not very close to 1, then $\alpha_*$ is of order of $\alpha_0$ and   larger than it. It should be noted that before the scale factor reaches $\alpha_*$ the solution crosses the line $\Gamma_-$. If at this point the control function $\chi$ vanishes, the solution can slip to the subcritical regime.
In the Appendix it is shown that such a solution for $\chi$ exists.
In such a case the solution for $\alpha$ leaves its supercritical phase and one gets an expanding Friedmann-Robertson-Walker universe filled with thermal radiation.

\section{A special case: Einstein constraint}\label{Sec7}

In the previous discussion we assumed that the parameter $\mu$ was positive. Let us discuss the supercritical solutions in the limiting case where this parameter tends to zero.
Using Eq.(\ref{zpq}), we rewrite Eq.(\ref{aa0}) in the form
\be
\ln(\alpha/\alpha_0)={\mu\over 2(1-\mu)}\left[ \ln \left( {1+\mu\over 2}\right) +\ln(1-\bq)\right]\, .
\ee
For  $\mu\ll 1$ one has
\be
\ln(\alpha/\alpha_0)={\mu\over 2}\ln\Big(\frac{1-\bq}{2}\Big)+O(\mu^2)\, .
\ee
The supercritical evolution starts at the point where $\bq=\lm=-(1-\mu)/(1+\mu)$ and continues its motion along the constraint line $C$ until it reaches a bounce point in a close vicinity of a point $\bq=1$. During practically the entirety of this evolution the ratio $a/a_0$ is of the order of 1. Essential change of the scale factor $\alpha$ occurs only when $\bar{q}$  becomes close to 1, so that
\be
1-\bq \sim \exp(-2/\mu)\, .
\ee

If we put $\mu=0$ directly into Eqs.(\ref{TC})--(\ref{SC}), we get
\be
{\cal Z}={\cal G}=p\hh \hat{\cal Z}=\hat{\cal G}={1\over 3}(p+2q)\, .
\ee
This means that such a limiting constraint is equivalent to putting restrictions on the eigenvalues of the Einstein tensor $\ts{G}$.
We call these restrictions the Einstein constraint.
The temporal Einstein constraint is $p={\Lambda}/3=$const.
The conservation law (\ref{hatG}) implies that the spatial constraint $\hat{\cal G}={\Lambda}$ is satisfied. The constraint line $C$ is vertical so that $\lp=\lm$. In the limit $\mu\to 0$ the parameter $\bq$ ``jumps" along this line from $-1$ to $1$. The solution of the constraint equation
\be
\alpha'^2-\alpha^2=1
\ee
is
\be
\alpha=\cosh(\tau)\, .
\ee
This is a de Sitter solution. This supercritical solution begins at $\tau=\tau<0$ where
\be
\cosh(\tau_0)=\alpha_0=\Big( \frac{\kappa C\lm}{3}\Big)^{\frac{1}{4}}\, .
\ee
After a bounce at the moment $\tau=0$ the universe begins to expand.


\section{Quadratic-in-curvature constraints}\label{Sec8}

\subsection{General remarks}

In our discussion of the linear-in-curvature constraints we imposed restrictions on the eigenvalues of the linear combinations of the Ricci tensor and the diagonal tensor proportional to the Ricci scalar. Let us now discuss a more general approach where the constraints are composed of functions of scalar invariants constructed from the Ricci tensor\footnote{We still do not consider invariants that contain covariant derivatives of this object.}. The corresponding constraint can be written in the form
\be\n{ffpq}
f(p,q,{\Lambda})=0\, .
\ee
This equation establishes a relation between the quantities $p$ and $q$, defined by Eq.(\ref{defpq}), and determines a corresponding constraint line $C$ in the $(p,q)$ plane.
Let us discuss some  general properties of such constraints.
Let us assume that ${\partial f\over \partial q}\ne 0$, so that the equation for the curve $C$  (at least over some its interval) can be written in the form $q=q(p)$. This is nothing but a second-order (nonlinear) equation which is resolved with respect to the second derivative,
\be\n{AAAA}
\ddot{a}=A(a,\dot{a})\, ,
\ee
where the function $A(a,\dot{a})$ is determined by the constraint equation (\ref{ffpq}).
It may happen that this nonlinear equation has a singular point at which the solution terminates. In such a case the corresponding constraint is singular.

To illustrate this let us use the relation
\be\n{dotp}
\dot{p}=2{\dot{a}\over a}(q-p)\, ,
\ee
which directly follows from  Eq.(\ref{defpq}) and which is equivalent to Eq.(\ref{AAAA}).

\begin{figure}[!hbt]
    \centering
    \vspace{10pt}
    \includegraphics[width=0.25 \textwidth]{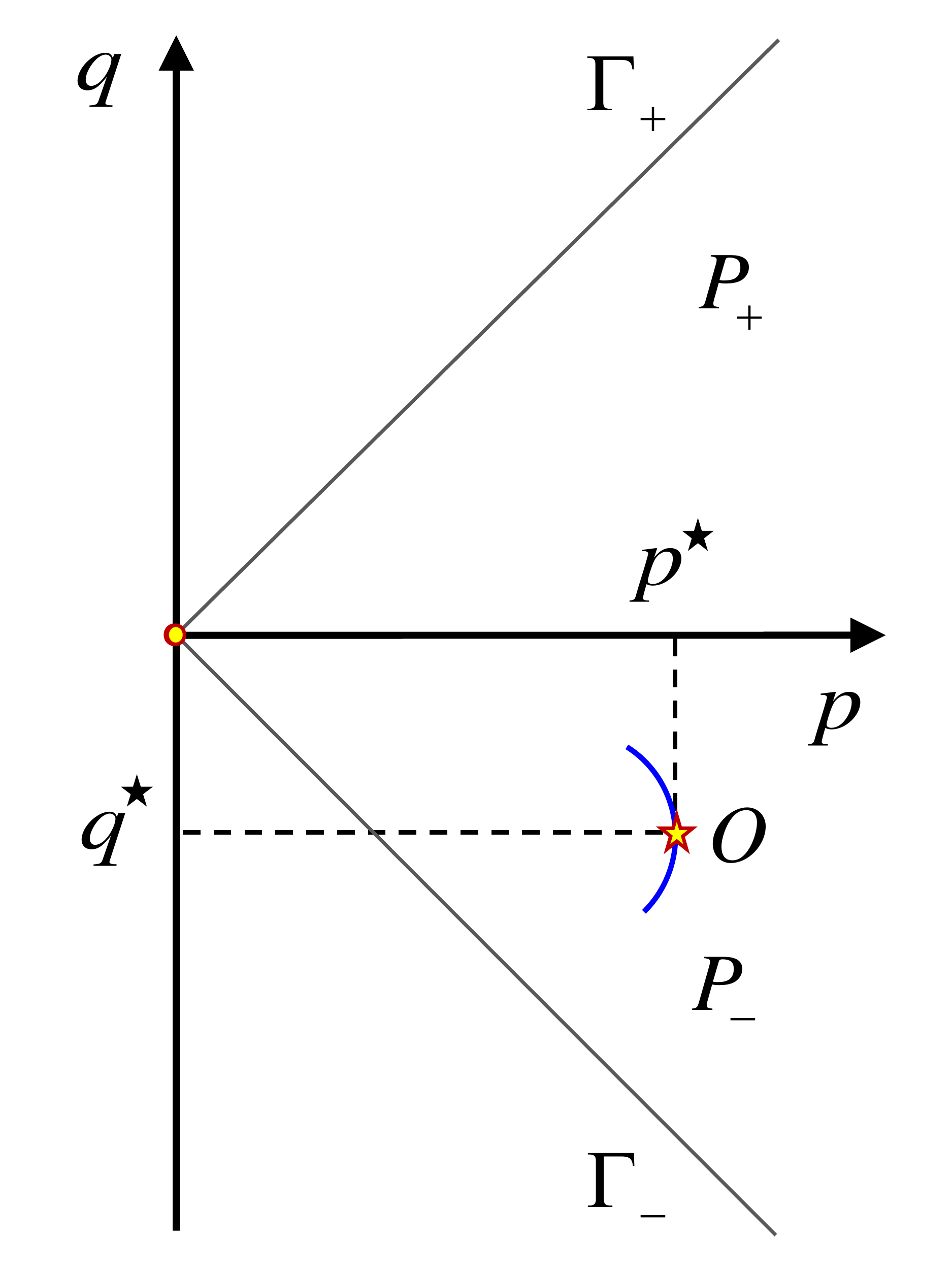}
    \caption{A curvature constraint is singular if the line $C$ representing it in the $p-q$ plane has a point $O$ in the $P_-$ domain where $p$ has a local maximum.}
    \label{Fig_pq_b}
\end{figure}

As earlier, we consider the evolution of a radiation-dominated universe at the state of contraction which is represented (see Fig.\,\ref{pq_0}) by the interval of line $\Gamma_-$ where $q=-p$. It starts at some point 1 and continues until it meats the constraint line $C$ at point 0. After this, the solution becomes supercritical and moves along the constraint line $C$. Since initially the universe contracts, $\dot{a}<0$ at point 0. Let us assume that in its further motion along the constraint a point representing the universe enters the $P_-$ domain (see Fig.\,\ref{Fig_pq_b}).  Since $q$ is negative there  $\dot{a}$ can only decrease and hence remains negative. A turning point of $a(t)$, if it exists, can only be located in the  domain $P_+$ where $q>0$.

\begin{figure}[!hbt]
    \centering
    \vspace{10pt}
    \includegraphics[width=0.25 \textwidth]{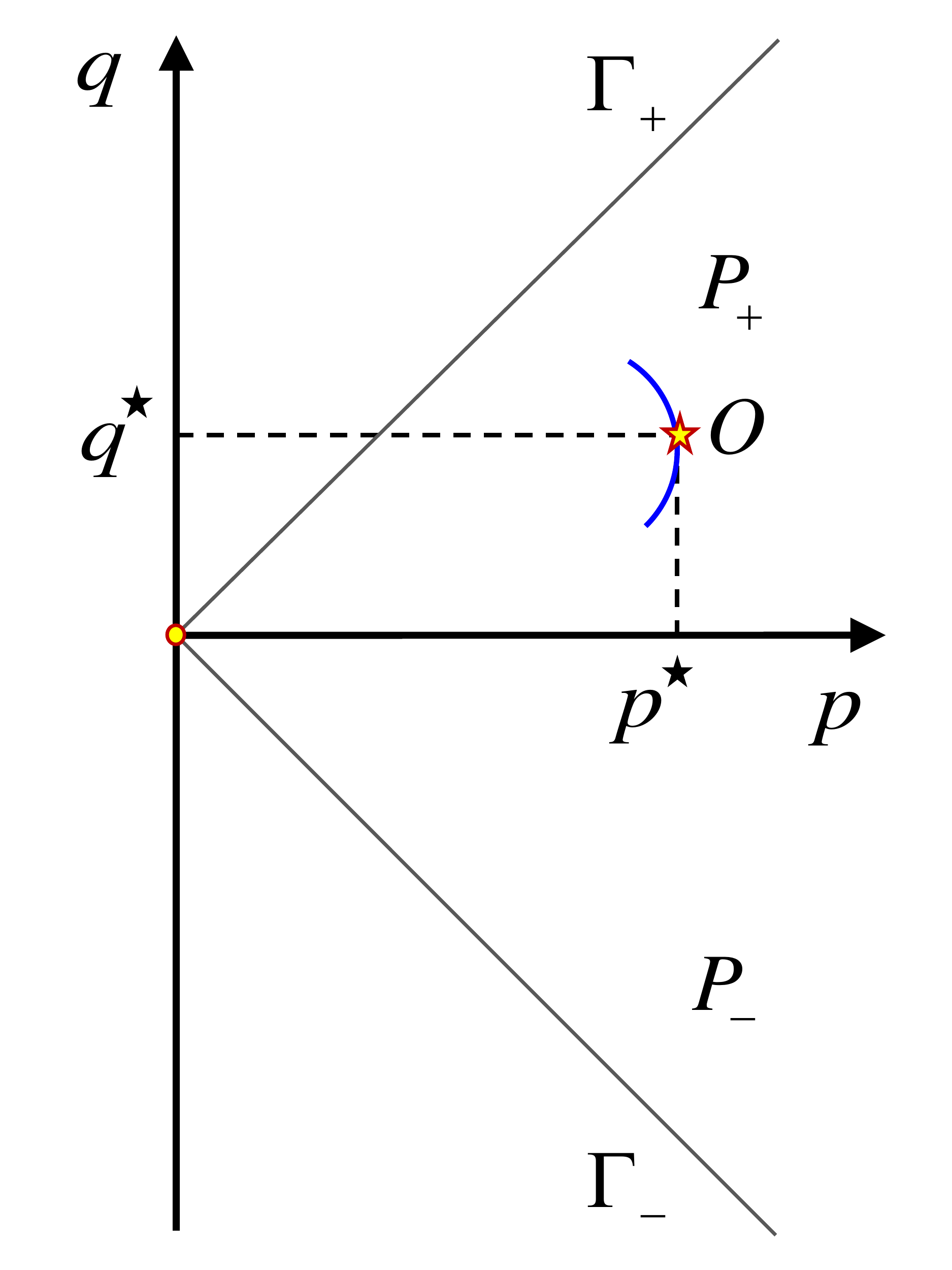}
    \caption{A curvature constraint is singular if the line $C$ representing it in the $p-q$ plane has a point $O$ in the $P_+$ domain where $\dot{a}$ is negative and $p$ has a local maximum. For $\dot{a}|_\ins{O}>0$, such a point cannot be reached by a supercritical solution.}
    \label{Fig_pq_a}
\end{figure}

Let us assume that the constraint equation does not allow the parameter $p$ to be bigger than $p^\star$ and a solution of the equation $f(p,q)=0$ near the point  $O$ with coordinates $(p^\star,q^\star)$ has the form shown in Figs.\,\ref{Fig_pq_b} and \ref{Fig_pq_a}.  In the $P_-$ domain $q-p<0$ and $\dot{a}<0$. Equation (\ref{dotp}) shows that $\dot{p}$ is positive there. Thus, in the vicinity of the point $O$ in the $P_-$ domain a point on the constraint curve representing a solution with the increasing time $t$ moves towards the point $O$ (see Fig.\,\ref{Fig_pq_b}).  A solution cannot be continued  beyond this point. The point $O$ itself is a singular point of the nonlinear second-order ordinary differential equation (\ref{AAAA}). Such a constraint is singular.

Consider a constraint that has a point  $O$ with the maximal value of $p$ located in the $P_+$ domain (see Fig.~\ref{Fig_pq_a}). If $\dot{a}$ is negative at  $O$, then using the above given arguments one can conclude that such a constraint is singular. Let us assume now that $\dot{a}$ at  $O$ is positive. Then, a point representing a solution moves away from  $O$ while $p$ is decreasing.
This means that if the motion along the constraint starts at point  $0$ on $\Gamma_-$, then the solution cannot reach the point $O$. This happens because before the solution reaches  $O$ where $\dot{a}>0$ it first reaches a point on the constraint line $C$ where $\dot{a}=0$. This is a turning point of the solution. At this point $a(t)$ reaches its minimal value.  After this the scale factor $a(t)$ increases and a point representing the solution moves back along the constraint curve with decreasing parameter $p$. In other words a point of the constraint  $O$ where $\dot{a}>0$ is not dangerous and the supercritical solution never reaches it.

In what follows we shall not consider singular constraints that do not allow a complete description of the evolution  of the universe. Let us note that a ``natural" quadratic-in-curvature constraint in which one restricts the Kretschmann invariant belongs to a class of singular constraints. This can be easily seen since the corresponding constraint function is $f=p^2+q^2-{\Lambda}$, and $p$ reaches its maximum when  $q=0$.
We shall focus on nonsingular constraints.  We shall demonstrate that for a wide class of quadratic-in-curvature constraints there exists a turning point of $a(t)$ located in $P_+$ domain, which for a ``large" initial size of the scale factor is always very close to the line $\Gamma_+$ where $q=p$.

\subsection{Quadratic-in-curvature constraints}

\begin{figure}[!hbt]
    \centering
    \vspace{10pt}
    \includegraphics[width=0.38 \textwidth]{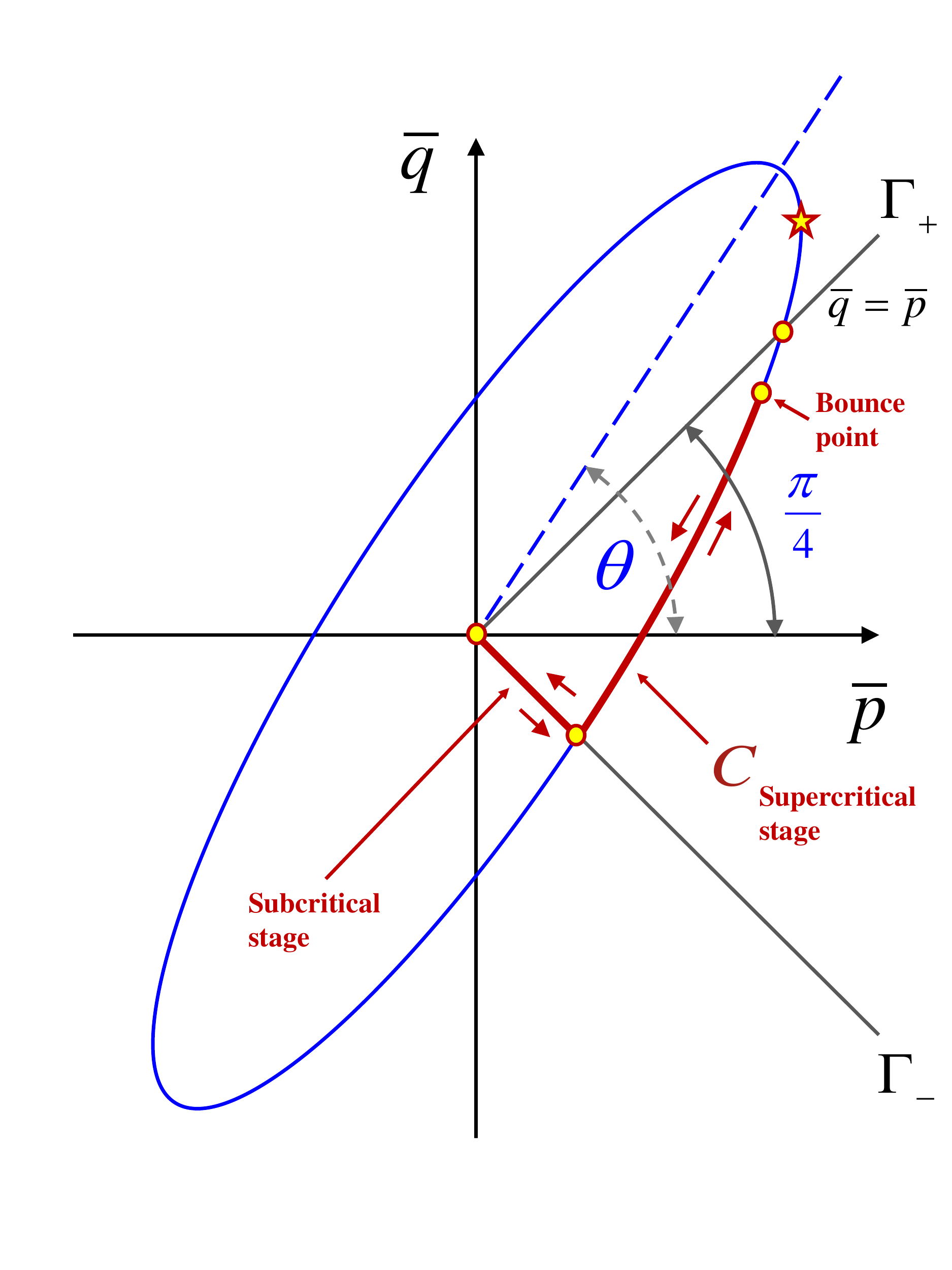}
    \caption{Quadratic constraint in $\bar{p}-\bar{q}$ plane. The red line from the center to the ellipse describes evolution during the radiation-dominated stage. The red arc along the ellipse describes evolution along the constraint untill the point $\bar{p}=\bar{q}$. This line $\bar{p}=\bar{q}$ corresponds to the de Sitter spacetimes. The point $\star$ with $[\bar{p}^\star,\bar{q}^\star]$ is the point where the tangent line to the ellipse is vertical $d\bar{p}/d\bar{q}=0$.}
    \label{Fig1}
\end{figure}

Let us now discuss a limiting curvature gravity model with quadratic-in-curvature constraints. We
denote
\be
\rho={1\over 6} R=p+q\hh  \sigma^2 ={1\over 3} S_{\mu\nu} S^{\mu\nu}=(p-q)^2\, .
\ee
The most general square-in-curvature expression can be written in the form
\be
Z=c_{SS} \sigma^2+ c_{SR} \sigma\rho+c_{RR} \rho^2\, .
\ee
As it will be explained later it is sufficient to use this constraint in the domain below $\Gamma_+$  where it can be written in the form
\ba
&Z=(c_{SS}+ c_{SR}+c_{RR}) p^2+ 2(c_{RR}-c_{SS})p q ,\\
&+(c_{SS} - c_{SR}+c_{RR}) q^2 .
\ea
The equation
\be \n{ZZLL}
Z={\Lambda}
\ee
determines  a second-order curve $C$ in the $(p,q)$ plane. We assume that this curve is an ellipse. The general ellipse can be parametrized by its two semiaxes ${A}$ and ${B}\le A$, and the angle $\theta$ between the large semimajor axis and coordinate axis $p$. In this parametrization  its equation is
\ba\label{ellipse1}
\Big(\frac{\cos^2\theta}{{A}^2}+\frac{\sin^2\theta}{{B}^2}\Big){p}^2
+\Big(\frac{\sin^2\theta}{{A}^2}+\frac{\cos^2\theta}{{B}^2}\Big){q}^2\\
+2\sin\theta\cos\theta\Big(\frac{1}{{A}^2}-\frac{1}{{B}^2}\Big){p}{q}=1 .
\ea
The coefficients $A$ and $B$ and the angle $\theta$ can be expressed in terms of the coefficients $c_{SS}$, $c_{SR}$, and $c_{RR}$ and ${\Lambda}$. In these variables the restriction on the curvature (\ref{ZZLL}) implies a restriction on the size of the ellipse and, in particular, on the ``length" of its major semiaxis $A$. A relation between the limiting curvature ${\Lambda}$ and $A$ can be easily found. Instead of this it is more convenient to choose from the very beginning the scale defined by $A$ as a limiting curvature parameter and to  use $A$ in order to introduce dimensionless quantities that describe our system. Namely, we set
\ba\label{barpq}
p={A}\bar{p}\hhh
q={A}\bar{q}\hhh a={\alpha\over \sqrt{A}} \hhh
t={\tau\over \sqrt{A}} \, .
\ea
We also denote
\ba
\gamma=\frac{{B}^2}{{A}^2}\hh 0\le\gamma\le 1\, .
\ea
Then the constraint equation (\ref{ellipse1}) takes the form
\ba\label{barconstraint}
&\bar{p}^2+\gamma\bar{q}^2\\
&-(1-\gamma)\big[(\bar{p}^2-\bar{q}^2)\cos^2\theta+2\bar{p}\bar{q}\cos\theta\sin\theta\big]=\gamma .
\ea

At the moment  when the radiation-dominated Friedmann stage matches the evolution along the constraint,
we have
\ba\label{plambda}
\bar{p}_{0}=-\bar{q}_{0}=\frac{\kappa C}{\Lambda a^4(t_{0})} \equiv {\lambda_{-}} .
\ea
Using this initial condition and the constraint \eq{barconstraint} we get the relation between $\lm$ and the parameters $\gamma$ and $\theta$
\ba\label{constraint0}
\lambda_{-}^2={\gamma \over (1+\gamma)+(1-\gamma)\sin2\theta}\, .
\ea

The point $\star$  where $dp/dq=0$ (see Fig.~\ref{Fig1}) has coordinates $(\bar{p}^\star,\bar{q}^\star)$
 \ba
&\bar{p}^\star=\sqrt{\cos^2\theta+\gamma\sin^2\theta} ,\\
&\bar{q}^\star=\frac{(1-\gamma)\sin\theta\cos\theta}{\sqrt{\cos^2\theta+\gamma\sin^2\theta} }.\\
\ea
We impose a condition $\bar{q}^\star>\bar{p}^\star$, that is the point $\star$ is located above $\Gamma_+$.
This is possible if
 \ba
0<\gamma< 3-2\sqrt{2}=(\sqrt{2}-1)^2.
 \ea
 For $\gamma=0$ the angle $\theta^\star$ at which the point $(\bar{p}^\star,\bar{q}^\star)$ lies on $\Gamma_+$ is
 \ba
 \theta^\star=\frac{\pi}{4}.
 \ea
For $\gamma=3-2\sqrt{2}$ the angle $\theta^\star$  is equal to
 \ba
 \theta^\star=\frac{3\pi}{8}=\arctan(1+\sqrt{2}).
 \ea
When  $0<\gamma\le 3-2\sqrt{2}$, the range of $\theta$ such that $q^\star>p^\star$ is
\ba
\theta_\ins{min}<\theta<\theta_\ins{max},
\ea
\ba\label{thetamin}
\theta_\ins{min}=\arctan\Big[\frac{1-\gamma-\sqrt{1-6\gamma+\gamma^2}}{2\gamma}\Big]\, ,\\
\theta_\ins{max}=\arctan\Big[\frac{1-\gamma+\sqrt{1-6\gamma+\gamma^2}}{2\gamma}\Big] \, .
\ea

The ellipse intersects $\Gamma_+$ at a point $\bar{q}=\bar{p}=\lambda_{+}$ where
\ba
\lambda_{+}^2=\frac{\gamma}{1+\gamma-(1-\gamma)\sin 2\theta}.
\ea
Note that for a fixed $\gamma$, $\lambda_{+}^2$ as a function of $\theta$ gets its maximum and minimum values at $\theta_\ins{max}$ and $\theta_\ins{min}$, respectively,
\ba
\lambda_{+}^2\big|_\ins{min}=\frac{\gamma(3\gamma-1
+\sqrt{1-6\gamma+\gamma^2})}{\gamma^2+4\gamma-1+(1-\gamma)\sqrt{1-6\gamma+\gamma^2}} ,
\ea
\ba
\lambda_{+}^2\big|_\ins{max}=\frac{\gamma(3\gamma-1
-\sqrt{1-6\gamma+\gamma^2})}{\gamma^2+4\gamma-1-(1-\gamma)\sqrt{1-6\gamma+\gamma^2}} .
\ea
For small $\gamma$,
\be
\lambda_{+}^2\big|_\ins{min}\simeq\gamma\hh
\lambda_{+}^2\big|_\ins{max}=\frac{1}{2}(1-\gamma) \,  .
\ee

\section{Evolution along the constraint and Big Bounce}\label{Sec9}

Let us now discuss the evolution of the universe in the supercritical regime for the quadratic constraint described in the previous section. A point representing the unverse in the $(\bar{p},\bar{q})$ plane starts its motion at $\bar{p}=\bar{q}=\lambda_-$ and moves with increasing $\bar{p}$ along the ellipse where $\bar{q}=\bar{q}(\bar{p})$. We now demonstrate that this monotonic motion continues until the point reaches the vicinity of  $\bar{p}=\bar{q}=\lambda_+$ where the scale factor $\alpha(\tau)$ has a turning point. For this purpose, we again use the following the relation, which follows from the definition of the quantities $\bar{p}$ and $\bar{q}$
\ba\label{barqp}
\frac{1}{2}\frac{d\bar{p}}{d\ln \alpha}=\bar{q}-\bar{p} .
\ea
It gives
\be\n{aaPP}
\alpha(\bar{p})=\alpha_0 \exp( -F)\hh
F=\frac{1}{2}\int_{\lambda_-}^{\bar{p}} {d\bar{p}\over \bar{p}-\bar{q}(\bar{p})}\, .
\ee
Here $\alpha_0$ is the dimensional value of the scale function $\alpha$ at the beginning of the supercritical evolution, that is at $\bar{p}=\lambda_-$. According to our assumption $\alpha_0\gg 1$.

In the next section we show that for the quadratic-in-curvature constraint the integral in Eq.(\ref{aaPP}) can be calculated explicitly. Now we demonstrate that the general form of the relation Eq.(\ref{aaPP}) allows one to prove that the supercritical solution always has a bounce. Let us note that  the integrand in the expression for $F$ is positive
 in the domain below $\Gamma_+$ where $\bar{p}<\lambda_+$ and $F$ is a monotonically increasing function of $\bar{p}$ which is logarithmically divergent at $\bar{p}=\lambda_+$.
One can use Eq.(\ref{aaPP}) to find $\bar{p}$ as a function of $\alpha$.
Using the definition of $\bar{p}$ one has
\be \n{ttt}
(\alpha ')^2=\bar{p}\alpha^2 -1\, .
\ee
Substituting $\bar{p}(\alpha)$ into this relation one obtains an equation that determines the evolution of the scale factor $\alpha$ as a function of time $\tau$.
If this function has a minimum $\alpha_\ins{b}$, then the following condition should be satisfied:
\be\n{turn}
{1\over \bar{p}_\ins{b}} \exp (2F_\ins{b})=\alpha_0^2 \, .
\ee
For $\lm\lesssim\lp$  the parameter $\bar{p}$ is  of order of one.  Since $F$ grows infinitely near $\bar{p}=\lambda_+$, Eq.(\ref{turn}) for large $\alpha_0$ always has a solution $\bar{p}_b$ which is located near $\lambda_+$.
This solution  determines the size of the universe $\alpha_b$ at the turning point.

The parameter $\alpha_b$  can be estimated as follows.
Near $\lambda_+$ one can write
\ba\label{deltapq}
\bar{p}=\lambda_{+}(1+\Delta p) \hh \bar{q}=\lambda_{+}(1+\Delta q)\, .
\ea
Using the ellipse equation, one finds
\ba\label{mup}
\Delta p-\Delta q=\mu \Delta p\, ,
\ea
where
\ba\label{mu}
\mu=-4\frac{1+\gamma-(1-\gamma)\sin 2\theta}{1+\gamma+(1-\gamma)(\cos 2\theta-\sin 2\theta)} .
\ea
In the range  $0<\gamma\le 3-2\sqrt{2}$ and
$
\theta_\ins{min}<\theta<\theta_\ins{max},
$ the parameter $\mu>0$.

If we denote by $\Delta p_b<0$ a position of the turning point, then
\ba
F_b=\frac{1}{2}\int_{\lambda_-}^{\lambda_+(1+\Delta p_b)} {d\bar{p}\over \bar{p}-\bar{q}(\bar{p})}\, .
\ea
The main contribution to this integral comes from the vicinity of its upper limit. This gives
\ba\label{N0}
F_b\approx -{1\over 2\mu}\ln |\Delta p_b|\, .
\ea
Equation (\ref{turn}) implies
\ba\n{DDTP}
\Delta p_b\approx - (\lambda_+ \alpha_0^2)^{-\mu}\, .
\ea
In the turning point
\be
\alpha_b={1\over \sqrt{p_b}}\, .
\ee
Then, using Eqs.(\ref{deltapq}) and  (\ref{DDTP}), one finds
\be
\alpha_b \approx {1\over \sqrt{\lp}}\left( 1+{1\over 2}(\lp \alpha_0^2)^{-\mu}\right)\, .
\ee
Hence at the turning point $\alpha_b$ is always larger than $\lp^{-1/2}$ and for large $\alpha_0$ its deflection from this value is  small. The existence of the turning point means that the universe has a bounce where its contraction is changed to the expansion.

\section{Exact solution}\label{Sec10}

Let us demonstrate that for the quadratic-in-curvature constraint one can obtain an explicit expression relating $\bp$ and $\alpha$. For this purpose we again use the equation
\be \label{barqp}
\bar{q}=\bar{p}+\frac{1}{2}\frac{d\bar{p}}{d\ln \alpha} .
\ee
Let us use Eq.\eq{barconstraint} to express $\bar{q}$ in terms of $\bar{p}$
\ba\n{qw}
\bar{q}=\frac{u \bar{p}-v\sqrt{w-\bar{p}^2}}{w}\, ,
\ea
where $u,v,w$ are the following constants
\ba
&u=(1-\gamma)\sin\theta\cos\theta \, ,\\
&v=\sqrt{\gamma} \, ,\\
&w=(1-\gamma)\cos^2\theta+\gamma  \, .
\ea
After substituting Eq.(\ref{qw}) into Eq.(\ref{barqp}) one obtains the first-order differential equation that determines $\alpha$ as a function of $\bp$. A solution of this equation is
\ba\label{lna}
\ln \Big(\frac{\alpha}{\alpha_{0}}\Big)&=\frac{w}{2[v^2+(u-w)^2]}\\
&\Big\{
-v\Big[\arcsin\Big(\frac{\bar{p}}{\sqrt{w}}\Big)-\arcsin\Big(\frac{\lm}{\sqrt{w}}\Big)\Big]\\
&+(u-w)\ln\Big|\frac{\bar{p}(u-w)-v\sqrt{w-\bar{p}^2}}{\lm(u-w)-v\sqrt{w-\lm^2}}\Big|
\Big\} .
\ea
The integration constant was fixed by the matching condition at the moment when this supercritical solutions starts and where  $\bar{p}=\lm$,
\ba
{\lambda_{-}}=\frac{\bar{\kappa}C}{3\alpha_{0}^4} .
\ea
Equation (\ref{lna}) defines $\bp$ as a function of $\alpha$. After substituting this function in (\ref{ttt}) one can find the time dependence of the scale factor $\alpha(\tau)$.

At the point $\bar{p}=\bar{q}=\lambda_+$ the argument of the logarithm in Eq.\eq{lna} vanishes.
Near this point, let $\bar{p}=\lambda_+(1+\Delta p)$. Then, the leading asymptotic of Eq.\eq{lna} at small $|\Delta p| \ll 1$ is
\ba\label{lna1}
\Delta p \simeq - \Big(\frac{\alpha}{\alpha_{0}}\Big)^{2\mu}\, ,
\ea
where
\ba
\mu &=\frac{2[v^2+(u-w)^2]}{w(u-w)}\\
&=-4\frac{1+\gamma-(1-\gamma)\sin(2\theta)}
{1+\gamma-(1-\gamma)(\sin(2\theta)-\cos(2\theta))}.
\ea
For every fixed $\gamma\in[0,3-2\sqrt{2}]$ the constant $\mu$ is positive in the range $\theta_\ins{min}<\theta<\theta_\ins{max}$ [see equation Eq.\eq{thetamin}[. For the turning point, where $\alpha=\alpha_b\approx \lp^{-1/2}$, Eq.(\ref{lna1}) correctly reproduces Eq.(\ref{DDTP}).

Let us summarize: the evolution of the universe for models with quadratic-in-curvature constraints is quite similar to the case of linear constraints. Namely, there exists a range of free parameters that specify a model for which these constraints are not singular. For such constraints there always exists a bounce and the supercritical solution is symmetric with respect to this moment of time. The size of the universe at the bounce is close to $\lp^{-1/2}$ and always larger than this parameter.
During this supercritical phase the contraction of the universe is replaced by its expansion. The results presented in the Appendix imply that there always exists a solution  of the gravity equations for the control function $\chi$ that is time symmetric with respect to the turning point. For this solution  at the moment when the point representing the expanding supercritical universe crosses $\Gamma_-$ it can slip to the subcritical regime.  This subcritical solution describes an expanding Friedmann universe filled with thermal radiation which contains the same entropy as the original collapsing world. The parameter  $N_\ind{b}=\ln \Big(\frac{\alpha_{0}}{\alpha_\ind{b}}\Big)$  for the expansion from the bounce point to the beginning of the Friedmann phase is nothing but the {\it e-fold number}
(for a review of the restrictions on the $e$-fold number in inflationary models from observations and their cosmological implications see, e.g., Ref.\cite{Mukhanov:2005sc} and references therein).

\section{General case}
\label{Sec11}

\begin{figure}[!hbt]
    \centering
    \vspace{10pt}
    \includegraphics[width=0.25 \textwidth]{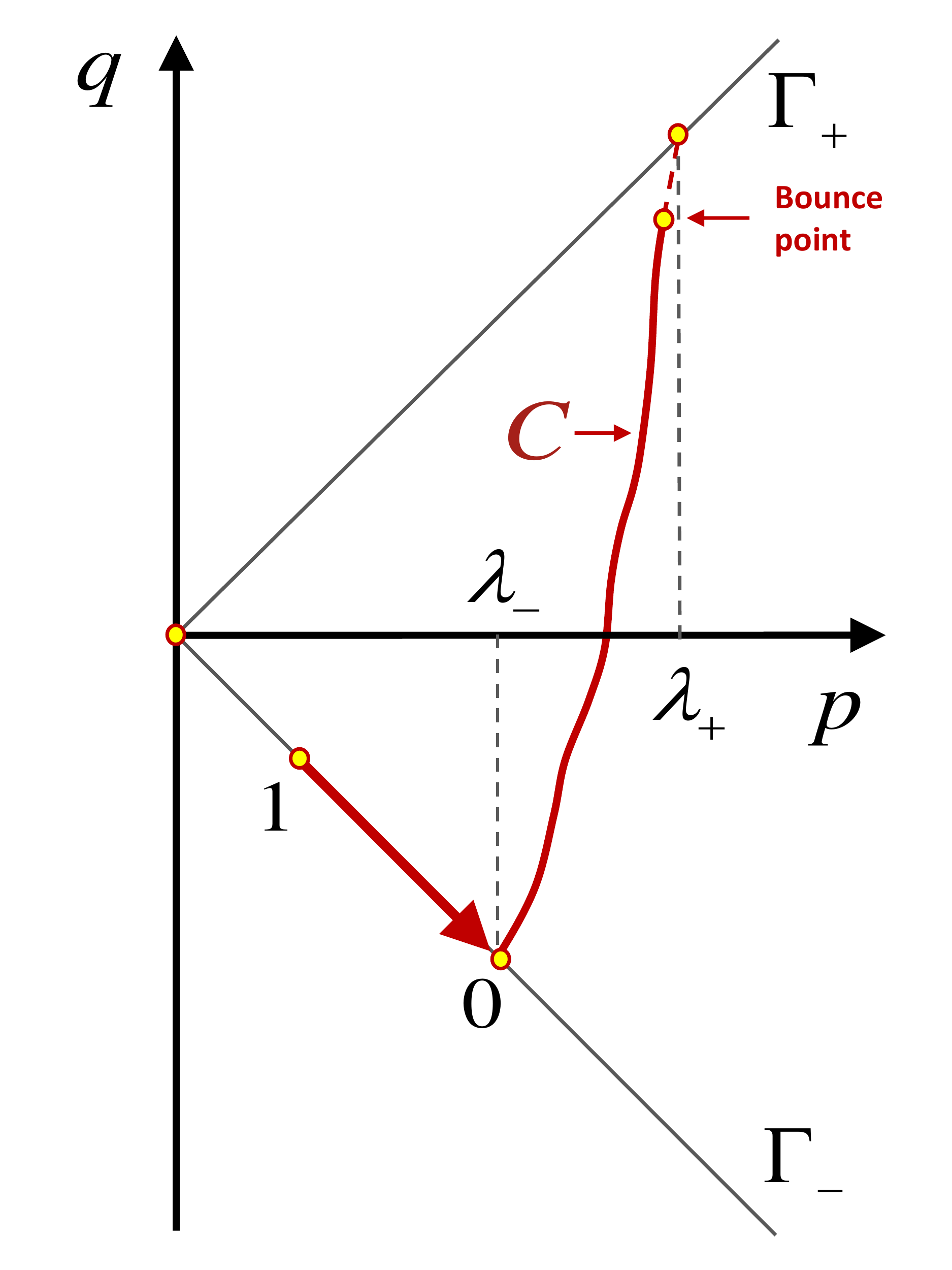}
    \caption{Behavior of a general constraint curve on ${p}-{q}$ plane.}
    \label{GEN}
\end{figure}

We have discussed cases of linear and quadratic-in-curvature constraints which admit rather complete analysis. In this section we demonstrate that under quite general assumptions many of the features of these models are also valid for a wider class of curvature constraints. Namely, we consider constraints constructed from the Ricci tensor invariants. As earlier, we do not include invariants containing derivatives of this tensor. Such a general constraint can be written in the form
\ba\label{ffff}
f(p,q, \Lambda)=0 .
\ea
Here $p$ and $q$ are defined by Eq.(\ref{defpq}) and $\Lambda$ is a parameter defining the limiting curvature. This constraint defines a line $C$ in the $(p,q)$ plane (see Fig.\,\ref{GEN}).

We make the following assumptions:
\begin{enumerate}
\item The constraint curve $C$ intersects both lines $\Gamma_-$ and $\Gamma_+$. We denote the coordinate $p$ at the intersection points by $\lm$ and $\lp$, respectively.
\item We assume that  ${\partial f\over \partial q}\ne 0$ on a segment of $C$ between $\Gamma_-$ and $\Gamma_+$, so that on the interval $p\in [\lm,\lp]$ one can express $q$ as a function of $p$ and $\Lambda$, $q=q(p,\Lambda)$.
\item This function on the interval $p\in [\lm,\lp]$ obeys the condition ${dq\over dp}>1$.
\end{enumerate}

The last condition implies that $\lp>\lm$. We denote by $\xi$ the minimal value of ${dq\over dp}$ on the interval $p\in [\lm,\lp]$. Then one has
\be
{\lp\over \lm} >{\xi+1\over \xi-1}\, .
\ee
We assume that $\xi$ is not very close to 1, so that the parameters $\lp$ and $\lm$ are of the same order.

One can use Eq.(\ref{App}) for the scale factor evolution for a supercritical solution with the constraint (\ref{ffff})
\be\n{GENa}
a=a_0 \exp(-F)\hh
F=\frac{1}{2}\int_{\lm}^p \frac{dp}{p-q(p)}\, .
\ee
Here $q(p)$ is defined by the constraint equation. We denote
\be
{1\over \mu}=\left.{dq\over dp}\right|_{p=\lp}\, .
\ee
Then, near $p=\lp$ one has
\be
p-q(p)\sim (\mu^{-1}-1)(\lp-p)\, .
\ee
Thus, the integrant in $F$ has a pole at $p=\lp$, so that this integral is logarithmically divergent at this point and
\be
\exp(F)\sim {1\over (\lp -p)^{(1-\mu)/\mu}}\, .
\ee
Using the same arguments as in the earlier discussion of linear and quadratic-in-curvature constraints, one can conclude that there exists a turning point where $a(t)$ has the minimal values. The scale factor at this point can be found by using Eq.(\ref{GENa}). After the bounce a point representing the supercritical solution moves back along the line $C$. To summarize, the supercritical evolution of the universe for the general constraint (\ref{ffff}) satisfying the conditions 1--3 is qualitatively the same as for linear and quadratic constraints. The  control function $\chi(t)$ for such a solution is discussed in the Appendix.

\section{Discussions}\label{Sec12}

In this paper we studied the evolution of an initially contracting isotropic homogeneous closed universe in the limiting curvature gravity theory. For this purpose, we modified a standard Einstein-Hilbert action by adding terms that restrict the curvature invariants. This was done in such a way that when the curvature is less than the critical one the evolution of the universe follows the standard (unconstrained) cosmological equations. We called this regime a subcritical one. For such a solution, the control function $\chi$ in the action vanishes.

After the spacetime curvature reaches its critical value a solution follows along the constraint and the control function $\chi$ becomes a nonvanishing function of time. The solution can leave its supercritical regime if the control function becomes zero. To make discussion more concrete we assumed that the contacting universe is initially filled with a thermal gas of radiation and a transition from sub- to supercritical regime occurs when the size of the universe is large in the following sense. If the critical value of the curvature is $\sim 1/\ell^2$, then we required that the size $a_0$ of the contracting universe at the moment when its curvature reaches this critical value obeys the condition $a_0/\ell \gg 1$.

There is a freedom in the choice of the term in the action which controls and restricts the growth of the curvature. We studied two types of constraints. The first class are constraints that are linear in  eigenvalues of the Ricci tensor. Such  constraints are represented by a straight line in $(p,q)$ plane. If this line crosses $\Gamma_{\pm}$ where $q=\pm p$ and $0<dp/dq<1$ for it, then we demonstrated that the evolution of the universe with such an inequality constraint is the following. After the contracting universe reaches the critical curvature and the evolution becomes supercritical, its acceleration parameter $q=\ddot{a}/a$ quite soon becomes positive. If the scale factor at the transition point $\alpha_0$ is large, then its further motion is very close to the motion of a contracting de Sitter universe. It has a bounce where the scale factor $a(t)$ has the minimal value $a_b$ close to $\ell$ and begins expanding. Both the scale function $a(t)$ and the control function $\chi(t)$ are symmetric under reflection with respect to the turning point. The function $\chi(t)$ becomes zero again when the size of the expanding universe becomes equal to $a_0$. After this point the solution leaves the constraint and it describes an expanding universe filled with thermal radiation. The entropy of this radiation is the same as that during the contraction phase.

The second class of constraints that we discussed in this paper are quadratic in curvature. We demonstrated that there exists a wide variety of such (nonsingular) constraints that guarantee that solutions are complete, that is, they do not break at a finite time. In the $(p,q)$ plane the constraint curves are ellipses with two parameters the angle $\theta$ characterizing the orientation of the ellipse, and the ratio of its semiminor and semimajor axes $\gamma$. The size of the major semiaxis characterizes the limiting curvature value. We showed that if $\theta$ and $\gamma$ obey some inequalities the corresponding constraint is nonsingular. The evolution of the universe for such nonsingular quadratic constraints is similar to the case of linear-in-curvature) constraints. After the universe reaches the  point where its curvature becomes critical, the solution evolves along the constraint. During this supercritical phase it reaches a point of bounce after which the scale function grows. The control function $\chi$ can become zero again at this phase and the universe can leave its supercritical regime. After this one has an expanding universe filled with thermal radiation which follows the corresponding solution of the Einstein equations.

Let us emphasize that these results  allow a natural generalization. In Sec.\,\ref{Sec11} we demonstrated  that they can be easily extended  to the case when a constraint is not linear or quadratic in the  curvature  but is described by a quite general function of it.\footnote{Here we do not consider more general invariants constructed from the curvature and its derivatives.} The qualitative behavior of the supercritical solutions in such models remains qualitatively the same. The solutions predict a bouncing point, when the contracting universe transitions to expansion.

In our discussion we assumed that a contracting universe is filled with thermal radiation. This simplified our analysis at one point, where we calculated the value of the scale parameter $a_0$  at the moment of transition of the solution from the sub- to supercritical regime. The value of $a_0$ can be easily found for any other choice of the equations of state. This changes nothing in the further  supercritical evolution of the universe, which was the main point of our discussion. Another assumption was that our contracting universe is closed. The other two cases, $k=0$ and $k=-1$ where the universe is open , can be analyzed similarly. The main difference is that a supercritical solution for $a(t)$ does not have a turning point but it can reach zero value. However, this does not mean that one has a physical singularity at this point. The curvature invariants remain finite and bounded and the singularity of the solution is a reflection of  a ``bad" choice of the coordinates. The situation here is similar to the case of  de Sitter model when coordinates with open space slices are chosen.
One can expect that by using proper coordinates one can further trace the evolution of the supercritical solution. It would be interesting to study these cases in detail and to confirm (or disprove) that there is also a bounce for these universes.

Many gravity models have been proposed in the literature that describe an inflationary stage of the Universe. Some of these models involve either higher-order-in-curvature terms or higher derivatives, or both. As a consequence, these models are typically prone to instabilities \cite{Yoshida:2017swb}. Some of these instabilities are related to the presence of ghosts (see however the discussion in Ref.\cite{deOSalles:2018eon}). Complications with ghosts can be avoided in some versions of nonlocal higher-derivative theories of gravity, and cosmologically viable models admitting nonsingular bouncing solutions can be constructed \cite{Biswas:2010zk,Biswas:2012bp,Kumar:2020xsl}. The analysis of the stability of cosmological solutions is a nontrivial problem in both ghost-free higher-derivative theories of gravity and systems with constraints \cite{Yoshida:2017swb}.

In the models of limiting curvature gravity discussed in the present paper a set of pairs of Lagrange multipliers $\chi_i$ and $\zeta_i$ entering in a specific combination was introduced. As a result, as soon as some function of curvature invariants does not accede its limiting value, the gravity equations are exactly those of the pure Einstein theory. This means that during a subcritical stage all of the degrees of freedom and physical effects are exactly the same as in general relativity. No extra instabilities and ghosts appear.
During the supercritical  stage the metric evolution is governed by the constraints. Matching conditions provide us with the initial data for the evolution of the metric and the Lagrange multipliers. Further evolution is unambiguous and respects the property of limiting curvature. If one includes a constraint involving the Weyl tensor, then the growth of all relevant curvature invariants can be bounded even if instability modes appear. Of course, in more realistic models one has to constrain all kinds of curvatures and take into account anisotropic  and other deviations from the background geometry \cite{Yoshida:2017swb,Kumar:2021mgc}. The control fields $\chi$ are very special.
They identically vanish in the subcritical regime and this property allows one to obtain the uniquely specified initial conditions for them at the beginning of the supercritical regime.
In the latter case the control field obeys the linear inhomogeneous equation. The initial conditions and the inhomogeneous term completely fix the solution for $\chi$. The control fields do not bring extra degrees of freedom to the system. In this sense they are not dynamical.

There is a well-known generic problem of bouncing cosmological models: the growth of the anisotropy at the stage of contraction. Even if the anisotropy is initially small and it can be described as a perturbation of isotropic homogeneous space, its amplitude for a physically reasonable equation of state  grows fast so that during the contraction at some its stage the anisotropy would become to affect the dynamics of the contracting universe. One can expect that in the models with limiting curvature this anisotropy growth could be suppressed by a proper choice of the constraints that contain not only Ricci-tensor but also Weyl-tensor invariants.
When properly included, the corresponding constraint would not allow infinite anisotropy growth. It is interesting and important to check whether this is really so.

One might interpret the obtained results as follows. After the curvature of the  contracting universe reaches its maximal value the matter does not contribute to the growth of the curvature. Instead, the further growth of its stress-energy tensor is compensated by the generation of the control field $\chi$. After passing the bounce and reaching the point of slipping back to the subcritical phase, the hidden thermal radiation (with its entropy) simply reappears. In this sense, the thermal state of the inflating universe arises without an additional reheating. Certainly this and other interesting features of the bouncing cosmologies in the limiting curvature gravity models require further detailed analysis.

\appendix*

\section{Evolution of the control function $\chi$}\label{AppA}

Let us consider the general curvature constraint that was discussed in Sec.\,\ref{Sec11}. We now  discuss the
gravity equations that are obtained by variation of the action including this constraint over the metric. During the supercritical stage the metric evolution is governed by the constraint. The gravity equations, in fact, describe evolution of the control function $\chi$. As we discussed earlier there is only one independent equation which can be obtained by variation of the dimensionally reduced action over the lapse function $b$.
The constraint (\ref{ffff}) can be obtained from the action
\ba
&\mathbb{I}_{\chi}=2\pi^2\int \dd t\,a^3 b\, \mathbb{L}_{\chi},\\
&\mathbb{L}_{\chi}=\chi \big[f(p,q,\Lambda)+\zeta^2\big].
\ea
by its variation over the control function $\chi$. Let us emphasize that in order to derive a complete set of gravity equations one should substitute general expressions (\ref{ppqq}) for $p$ and $q$ which contain the gauge function $b$.

In what follows we assume that conditions 1--3 formulated in Sec.\,\ref{Sec11} are satisfied and one can use Eq.(\ref{GENa}) to find $p$ on the constraint line $C$ as a function of the scale parameter $a$.

Let us consider the gravity equation
\ba
&\frac{1}{v a^3}\frac{\delta [ \mathbb{I}_g+\mathbb{I}_{\chi}+\mathbb{I}_{m}]}{\delta b}\Big|_{b=1}=0
\ea
evaluated on the constraints \eq{ffff} and $\zeta=0$ becomes
\ba
\int \dd t \Big[\frac{\partial f}{\partial p} \frac{\delta p}{\delta b}
+\frac{\partial f}{\partial q}\frac{\delta q}{\delta b}\Big]\chi \Big|_{b=1}=\frac{1}{\kappa}{\cal G}-\frac{C}{a^4}  \, .
\ea

Taking into account that for any function $h(a)$
\ba
\int \dd t\, \frac{\delta p}{\delta b} h \Big|_{b=1}&=-2\frac{\dot{a}^2}{a^2} h
=-2\Big[p-\frac{1}{a^2}\Big] h,
\ea
\ba
\int \dd t\, \frac{\delta q}{\delta b} h&=-\Big[\frac{\ddot{a}}{a}+\frac{\dot{a}^2}{a^2}\Big]h+\frac{\dot{a}}{a}\dot{h}\\
&=-\Big[p+q-\frac{1}{a^2}\Big] h+\Big[p-\frac{1}{a^2}\Big]\frac{d h}{d \ln a} ,
\ea
we get  the first-order linear inhomogeneous differential equation
\ba\label{eqchi}
&\Big[p-\frac{1}{a^2}\Big] \frac{d \big(\frac{\partial f}{\partial q} \chi\big)}{d \ln a}-2\Big[p-\frac{1}{a^2}\Big] \frac{\partial f}{\partial p} \chi \\
&-\Big[p+q-\frac{1}{a^2}\Big] \frac{\partial f}{\partial q}\chi   =\frac{3p}{\kappa}-\frac{C}{a^4}\, .
\ea
This equation does not explicitly depend on time, but only on the parameter $a$. Written in this form it determines $\chi$ as a function of the scale factor $a$.

Since $\frac{\partial f}{\partial q}\neq 0$ on the interval $p\in[\lambda_{-},\lambda_{+}]$  we can redefine the $\chi$  function as
\ba
\omega=\frac{\partial f}{\partial q}\chi
\ea
and the zeros of the control functions $\chi$ and $\omega$ are the same.
Then, one can write the equation Eq.\eq{eqchi} in the form
\ba\label{eqomega}
\frac{d \omega}{d\ln a}-U(a)\,\omega= W(a) \, ,
\ea
where $y=\ln a$ , $p=p(a)$, and
\ba
U=1+\frac{q}{p-\frac{1}{a^2}}+2\frac{d q(p)}{d p},
\ea
\ba\label{W}
W=\frac{\frac{3p}{\kappa}-\frac{C}{a^4}}{p-\frac{1}{a^2}} .
\ea
At the moment $t_0$ we have $a(t_0)=a_0$ and $W(t_0)=0$ because the Einstein equations in a subcritical stage require
\ba\label{W0}
\frac{3p}{\kappa}-\frac{C}{a^4}=0
\ea
identically.

The solution for the control function $\omega$ reads
\ba\label{omega}
\omega=e^{\int_{y_0}^y \dd y \,U(y)}\Omega(y) ,
\ea
\ba\label{Omega}
\Omega(y)=\Omega_0+\int_{y_0}^y \dd y \,W(y) e^{-\int_{y_0}^y \dd z \,U(z)}  ,
\ea
where $y=\ln a$ and $y_0=\ln a_0$. At the matching point the control function vanishes. This condition fixes the integration constant $\Omega_0=0$.

The solution \eq{omega}-\eq{Omega} is finite for all $a$ between $a_0$ and the bounce radius $a_\ins{b}$. This fact is not evident and needs special analysis because both functions $U$ and $W$ have a pole near the bounce point. This happens because the function $p-\frac{1}{a^2}=\frac{\dot{a}^2}{a^2}$ in their denominators vanishes at the bounce, where $\dot{a}=0$. In order to prove that these poles do not lead to singularities for $\omega$ at $a_\ins{b}$, we analyze Eq.\eq{eqomega} in its vicinity. Let us analyze the asymptotic of this equation when $a\to a_\ins{b}$. Let $x=\ln( a/a_b)$ be a small parameter. Then, using the expansion near the turning point, one gets
\ba
p-\frac{1}{a^2}&=\frac{\dot{a}^2}{a^2}=\frac{\dot{a}^2}{a^2}\Big|_\ins{b}+\frac{a}{\dot{a}}\frac{d}{dt}\Big(\frac{\dot{a}^2}{a^2}\Big)\Big|_\ins{b} x +O(x^2)\\
&=2\frac{\ddot{a}}{a}\Big|_\ins{b} x +O(x^2)=2 q_\ins{b} x +O(x^2).
\ea
Here $q_\ins{b}=q(a_\ins{b},a_0,\Lambda)$ and  $p_\ins{b}=p(a_\ins{b},a_0,\Lambda)$. As soon as we know the evolution of the scale factor $a(t)$ along the constraint, we know $p_\ins{b}$ and  $q_\ins{b}$ and therefore can determine the evolution for the control functions $\omega$ and $\chi$.

For small $x$ one has
\ba
U=\frac{1}{2x}+O(1)\hh
W=-\frac{C_1}{2x}+O(1)\, ,
\ea
where the constant $C_1$ is defined by the asymptotic behavior of $W$ at the bounce point
\ba
C_1=\frac{3p_\ins{b}}{\kappa q_\ins{b}}\Big(\frac{\kappa C}{3}p_\ins{b}-1\Big)
\hh
\frac{\kappa C}{3}=a_0^4\lambda_{-} \, .
\ea
Then, the asymptotic of Eq.\eq{eqomega} near the turning point takes the form
\ba\label{eqomega2}
\frac{d \omega}{d x}-\frac{1}{2 x}\omega=-\frac{C_1}{2x}+O(1)\, .
\ea
Because at the bounce point $p_\ins{b}\sim q_\ins{b}\sim \lambda_{+}>0$ and  $a_0^2\lambda_{+}$ is a very big number, the constant $C_1$ is positive and  big too. Equation \eq{eqomega2} shows that
\ba
\omega=C_1+C_2\sqrt{|x|} +O(x).
\ea
The integration constant $C_2$  can be determined using the matching condition at $a_0$. If $t_\ins{b}$ is the moment of bounce, then near this point one has $a\approx a_\ins{b}[1+\frac{q_\ins{b}}{2}(t-t_\ins{b})^2]$. So that

\ba\n{omeg}
\omega\big|_{a\to a_\ins{b}}  \approx  C_1+C_2\sqrt{\frac{q_\ins{b}}{2}}\, |t-t_\ins{b}| .
\ea
Thus at the bounce point $\omega$ is finite and approaches its limiting value linearly in time. At the other moments all integrands in Eqs.\eq{omega}--\eq{Omega} are finite and the integrals are finite too. Let us note that for the solution (\ref{omeg}) the time derivative of $\omega$ at $t=t_\ins{b}$ has a jump. Such a jump is allowed by Eq.(\ref{eqchi}) for $\chi$ since at this point the coefficient of the term containing the derivative of $\omega$ vanishes.

Now let us prove that under the imposed conditions 1--3 of Sec.\,\ref{Sec11} the function $W$ is always negative,
provided the constraints satisfy the condition $d q(p)/d p>0$. Taking into account that $\frac{\kappa C}{3}=a_0^4\lambda_{-}$ we rewrite $W$ in the form
\ba
W=-\frac{3}{\kappa}\frac{a^2}{\dot{a}^2}\,Y \hh Y= \frac{a_0^4}{a^4}\lambda_{-}-p.
\ea
Using Eq.\eq{App}, we express $Y$ in terms of $p$ and $q(p)$,
\ba
Y(p)=\lambda_{-} e^{2\int_{\lm}^p \frac{dp}{p-q(p)}}-p\, .
\ea
From the matching condition \eq{W0} at the point $p=\lambda_{-}$ we have
\ba
Y\big|_{p=\lambda_{-}}=0 .
\ea
Its derivative
\ba
\frac{dY}{dp}=\frac{2Y+p+q}{p-q}
\ea
also vanishes at the matching point $p=\lambda_{-}$, $q=-\lambda_{-}$
\ba\label{dY}
\frac{dY}{dp}\Big|_{p=\lambda_{-}}=0\, .
\ea
The second derivative of $Y$  is
\ba\label{ddY}
\frac{d^2 Y}{dp^2}=2\frac{(Y+p)(1+\frac{dq(p)}{dp})}{(p-q)^2}\, .
\ea
Since $dq/dp$ is positive on the interval $p\in [\lm,\lp]$ the second derivative of $Y$ is positive at the matching point.
This means that in the vicinity of $\lm$ for $p>\lm$ both $Y$ and ${dY}/{dp}$ are positive. To prove that $Y(p)$ is positive on the whole interval  $(\lm,\lp)$ we assume the opposite. Namely, we assume that $Y$ becomes negative at some point on this interval. This means that there exists point a $p_1\in(\lm,\lp)$, where $Y$ vanishes again.
This may happen only if $Y$ reaches its maximum at $p_2\in (\lm,p_1)$. In this case ${dY}/{dp}|_{p_2}=0$ and, hence, there exists $p_3\in (\lm,p_2)$ where ${dY}/{dp}$ has a maximum. At this point ${d^2Y}/{dp^2}|_{p_3}=0$, ${dY}/{dp}|_{p_3}>0$, and $Y(p_3)>0$. This conclusion is in contradiction with Eq.\eq{ddY} evaluated at the point $p_3$.
Thus, our assumption that $Y$ may vanish inside the interval  $(\lm,\lp)$ leads to a contradiction and we must conclude that both $dY/dp$ and $Y$ are non-negative during the whole supercritical stage. Hence, $W$ is negative for $a<a_0$ and $W=0$ at $a=a_0$.

Using this property, one can show that $\omega$ is negative during the whole evolution along the constraint. The function $\chi$ differs from $\omega$ only by a factor $\partial f/\partial q$ and, hence, it does not vanish as well. This means that the control function $\chi$ does not vanish during a supercritical stage except for the initial and final matching points corresponding to the scale factor $a_0$.


\section*{Acknowledgments}

The authors thank the Natural Sciences and Engineering Research Council of Canada and the Killam Trust for their financial support. The authors are grateful to Andrei Frolov for stimulating discussions.



%

\end{document}